\documentclass{jpp}
\usepackage{graphicx}
\usepackage[dvipsname]{xcolor}
\usepackage{amsmath,amsfonts}
\usepackage{subfigure}
\usepackage{bm}
\usepackage[outdir=./]{epstopdf}
\usepackage[colorlinks=true, linkcolor=olive,citecolor=olive]{hyperref}

\shorttitle{Quasilinear theory of Brillouin resonances in rotating magnetized plasmas}
\shortauthor{J.-M. Rax, R. Gueroult and N. J. Fisch}

\title{Quasilinear theory of Brillouin resonances in rotating magnetized plasmas}

\author{J.-M. Rax\aff{1,2},  R. Gueroult\aff{3}
  \and N. J. Fisch\aff{4}}

\affiliation{\aff{1}Andlinger Center for Energy + the Environment, Princeton
University, Princeton, NJ 08540, USA
\aff{2}IJCLab, Universit\'{e} de Paris-Saclay, 91405 Orsay, France
\aff{3}LAPLACE, Universit\'{e} de Toulouse, CNRS, INPT, UPS, 31062 Toulouse, France
\aff{4}Department of Astrophysical Sciences, Princeton University, Princeton NJ 08540, USA}

\begin{document}

\maketitle

\begin{abstract}
Both spin and orbital angular momentum can be exchanged between a rotating wave and a rotating magnetized plasma. Through resonances the spin and orbital angular momentum of the wave can be coupled to both the cyclotron rotation and the drift rotation of the particles. It is however shown that the Landau and cyclotron resonance conditions which classically describe resonant energy-momentum exchange between waves and particles are no longer valid in a rotating magnetized plasma column. In this case a new resonance condition which involves a resonant matching between the wave frequency, the cyclotron frequency modified by inertial effects and the harmonics of the guiding center rotation is identified. A new quasilinear equation describing orbital and spin angular momentum exchanges through these new Brillouin resonances is then derived, and used to expose the wave driven radial current responsible for angular momentum absorption.
\end{abstract}

\section{Introduction}

Understanding how to sustain and control the angular momentum of a rotating magnetized plasma column is a central issue, both for applied and basic plasma physics. On the former, the first successful application of rotating non neutral plasmas was the magnetron microwave source theorized by Brillouin~\citep{Brillouin1945}. Since then, an important motivation for rotating plasma configurations has been and continue to be thermonuclear fusion~\citep{Lehnert1971}, both with rotating tokamaks~\citep{Rax2017,Ochs2017b} and rotating mirrors~\citep{Bekhtenev1980,Hassam1997,Fetterman2008,Fetterman2010,Teodorescu2010}. Besides fusion, rotating plasmas have also attracted attention for ion acceleration~\citep{Janes1965,Janes1965a,Janes1966} and
mass separation~\citep{Bonnevier1966,Krishnan1981,Prasad1987} as envisioned for instance for nuclear waste cleanup~\citep{Gueroult2015}, spent nuclear fuel
reprocessing~\citep{Gueroult2014a,Timofeev2014,Vorona2015,Dolgolenko2017} or rare earth element recycling~\citep{Gueroult2018a}.
On the latter, magnetised rotating plasma theory has been shown to be important
to understand pulsar dynamics and radiative transfer~\citep{Gueroult2019a}, rotation augmented gyrotropy~\citep{Gueroult2020} or image rotation (aka Faraday-Fresnel
effect) in plasmas~\citep{Rax2021}. The adiabatic
theory of angular momentum perturbation in rotating magnetized plasmas
also provides an interesting realization of a geometrical Berry type phase~\citep{Rax2019a}.

\begin{figure}
\begin{center}
\includegraphics[height = 6cm]{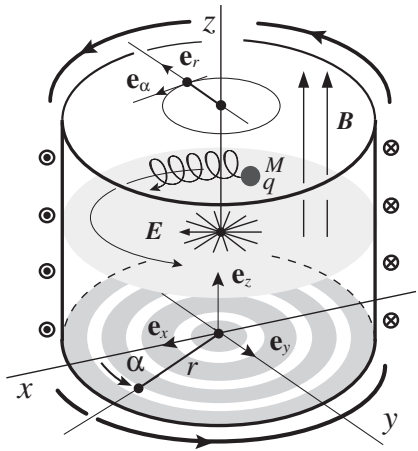}
\caption{Electric and magnetic fields configurations in a plasma rotating around the $z$ axis, $\alpha$ is the polar angle and $r$ the polar radius.}
\label{Fig:Fig1}
\end{center}
\end{figure}

Two fields configurations can sustain the steady state rotation of a magnetized plasma. One is the 
\textit{Hall configuration} with a radial magnetic field and an axial electric field. This is the configuration used notably in stationary plasmas thrusters. The other, illustrated in figure~\ref{Fig:Fig1}, is the \textit{Brillouin configuration} with an axial magnetic field and a radial electric field. This is the configuration used notably in mass separators~\citep{Gueroult2019} and homopolar devices~\citep{Barber1972}. In this study we will restrict our analysis to this last configuration and study the quasilinear theory of angular momentum exchange between waves and particles in a rotating Brillouin configuration.

Steady state angular momentum injection to compensate dissipation in a rotating cylindrical magnetized plasma column is usually envisioned through the use of concentric annular DC biased end-electrodes, as illustrated in gray in figure~\ref{Fig:Fig1}. This is the scheme originally proposed by Lehnert~\citep{Lehnert1970,Lehnert1973}, where electrodes are assumed to drive a DC radial
electric field perpendicular to the DC axial magnetic field. It has however been recently shown that field penetration through the sheath and along field lines imposes a number of constraints on the achievable electric field~\citep{Gueroult2019b,Poulos2019,Liziakin2020,Liziakin2021,Trotabas2022}. Meanwhile,  studies in the last decade on rotating mirrors~\citep{Fetterman2008,Fetterman2010} and rotating tokamaks~\citep{Rax2017,Ochs2017b} have underlined the potential to use electromagnetic
waves to drive plasma rotation through resonant wave-particle angular
momentum absorption, and to compensate for dissipative relaxation. Although very promising, these studies relied on a simple photon picture.

The standard tool to study resonant wave-particle interaction is the quasilinear theory. For an infinite homogeneous magnetized plasma at rest the quasilinear equation is well known~\citep{Rax2011}, and has proven particularly useful to evaluate energy absorption and current generation in thermonuclear plasmas~\citep{Fisch1978,Fisch1987}. The quasilinear equation for a cylindrical unmagnetized plasma at rest has also been derived~\citep{Kaufman1971}, whereas a generalised quasilinear theory for inhomogeneous plasma has recently been laid out~\citep{Dodin2022}. However, and while as mentioned above it appears to be key to important applications, the quasilinear theory for a rotating magnetized plasma has to our knowledge not been derived yet. In this paper we address this issue and derive the quasilinear equation for a rotating wave and a rotating magnetized plasma, which describes angular momentum absorption/emission within a kinetic framework, and use it to uncover the interplay between orbital angular momentum (OAM), spin angular momentum (SAM) and finite Larmor radius (FLR) effects.

This paper is organized as follows. We begin by recalling in Sec.~\ref{Sec:2} basic elements on wave and particle angular momentum. We then proceed to derive in Sec.~\ref{Sec:3} the Hamiltonian description of a magnetized rotating plasma sustained by a radial electric field, and to expose the physical and geometrical meaning of the angle/action variables used to describe the particle dynamics. These canonical angle-action variables are after that used to identify the relation between the canonical and the kinetic angular momentum in Sec.~\ref{Sec:4}, and to model as a perturbed Hamiltonian the effect of a rotating wave perturbation on the rotating particle dynamics in Sec.~\ref{Sec:5}. This formalism is then used to identify a new resonance condition in Sec.~\ref{Sec:6}, and to derive the quasilinear kinetic equation describing the time evolution of the distribution function in action space in Sec.~\ref{Sec:7}. This new kinetic equation is finally used to study OAM and SAM absorption in Sec.~\ref{Sec:8}, and to derive the expression of the wave driven radial current in Sec.~\ref{Sec:9}. Lastly, Sec.~\ref{Sec:10} concludes this study and summarizes the main findings.



\section{Wave and particle angular momentum}
\label{Sec:2}

In the following we use $\left( r,\alpha ,z\right) $ a set of cylindrical coordinates on a cylindrical basis $\left( \mathbf{e}_{r},\mathbf{e}_{\alpha },\mathbf{e}_{z}\right) $. The associated set of Cartesian
coordinates is $\left( x,y,z\right) $ on the Cartesian basis $\left( \mathbf{e}_{x},\mathbf{e}_{y},\mathbf{e}_{z}\right) $. The vertical axis along $z$ is the direction of the background static magnetic field $\mathbf{B=}B\mathbf{e}_{z}$ and the DC electric field $\mathbf{E}=E\mathbf{e}_{r}$ is along the radial direction. This is the configuration illustrated in figure~\ref{Fig:Fig1}.

Electromagnetic waves can carry both spin angular momentum - associated with right (R) and left (L) hand circular polarizations - and orbital angular momentum~\citep{Gough1986,Enk1994,Barnett1994,Goette2007,Barnett2017}. Plasma waves carrying orbital angular momentum has as a matter of fact been an active research topics in the last decade, both in unmagnetized~\citep{Mendonca2012,Chen2017,Bliokh2022} and magnetized~\citep{Shukla2012,Stenzel2015a,Stenzel2016a} plasmas. The vector field of such a wave has a helical phase front structure and can in general be written as 
\begin{equation}
\Real\left[\frac{\mathbf{e}_{x}\pm j\mathbf{e}_{y}}{\sqrt{2}}\mathcal{E}_{\pm
}(r)\exp j\left( n\alpha +\beta z-\omega t\right)\right].  \label{oam1}
\end{equation}
Here $\left( \mathbf{e}_{x}+j\mathbf{e}_{y}\right) \mathcal{E}_{+}$ is a R circularly polarized field while $\left( \mathbf{e}_{x}-j\mathbf{e}_{y}\right) \mathcal{E}_{-}$ is a L circularly polarized field, the electric field $\mathcal{E}(r)$ is the solution of the radial part of Maxwell's equations, $\omega $ is the wave frequency, $n\in \Bbb{Z}$ is the azimuthal mode number and $\beta \in \Bbb{R}$ is the axial wavevector. An observer located at a fixed point $\left( r_{0},\alpha _{0},z_{0}\right) $ and probing azimuthally the electric field amplitude $\left| \mathcal{E}_{\pm }(r_{0})\right| $ of the wave described by Eq.~(\ref{oam1}) will measure a field pattern $\left| \mathcal{E}_{\pm }(r_{0})\right| $ rotating at the angular velocity $d\alpha/dt$ = $\omega /n$. The formal identification of spin angular momentum (SAM) and orbital angular momentum (OAM) contents for the wave vector field given in Eq.~(\ref{oam1}), as well as the definition of the associated SAM and OAM operators $\widehat{\mathbf{S}}$ and $\widehat{\mathbf{L}}$, are discussed in Appendix~\ref{Sec:SAMOAM}.


Meanwhile, magnetized charged particles in axisymmetric fields can also carry both \textit{spin} angular momentum and orbital angular momentum. The former is associated with the cyclotron motion while the latter is associated with the guiding center motion around the $z$ axis.  Note that for classical particles, the separation of angular momentum into cyclotron SAM and drift OAM does not arise from a quantum analysis. It is a simple application of Koenig's theorem which states that the angular momentum of a system can be decomposed into an external orbital part and an internal part. This internal part is nowadays called \textit{spin} part for waves and magnetized charges, even within a classical framework. Note also that while the quantum SAM of the charged particles $\pm \hbar /2$ should in principle be considered along with the cyclotron SAM and the drift OAM, it will be neglected in this study as the plasma temperature is assumed to be far larger than $\hbar \omega _{c}$ which is of the order of $10^{-7}$~eV for protons and a magnetic field of one tesla. Consider now more specifically the plasma column shown in figure~\ref{Fig:Fig1} with a background axial magnetic field $\mathbf{B=}B\mathbf{e}_{z}$ and a radial electric field $\mathbf{E}=E\mathbf{e}_{r}$, which leads to a guiding center $E\times B$ rotation around the $z$ axis. An ion with charge $q$ and mass $M$ is described by (\textit{i}) its instantaneous position $\mathbf{r}$ = $\mathbf{R}_{G}$ + $\mathbf{\rho }_{L}$%
, where $\mathbf{R}_{G}$ is the guiding center position and $\mathbf{\rho }_{L}\left( t\right) $ the Larmor radius, and (\textit{ii}) its velocity $\mathbf{v}$ = $\mathbf{V}_{G}$ + $\mathbf{v}_{c}$, where $\mathbf{V}_{G}$ $\sim $ $\left( E/B\right) \mathbf{e}_{\alpha }$ is the guiding center drift velocity and $\mathbf{v}_{c}\left( t\right) $ = $\omega _{c}\mathbf{e}_{z}\times \mathbf{\rho }_{L}$ is the cyclotron velocity with $\omega _{c}=qB/M$ the ion cyclotron frequency. The instantaneous angular momentum is defined as $M\mathbf{r}\times \mathbf{v}$
and its average $\left\langle {}\right\rangle $ over the fast cyclotron motion is
\begin{equation}
M\left\langle \left( \mathbf{R}_{G}+\mathbf{\rho }_{L}\right) \times \left( 
\mathbf{V}_{G}+\mathbf{v}_{c}\right) \right\rangle \cdot \mathbf{e}%
_{z}=MR_{G}^{2}\varpi +M\rho _{L}^{2}\omega _{c}=L_{z}+S_{z}.
\label{sam22}
\end{equation}
where $\varpi $ = $\left( E/BR_{G}\right) $ is the angular $E\times B$ drift velocity. The OAM part of Eq.~(\ref{sam22}) is $L_{z}$ $=M_{G}\varpi $ with $M_{G}$ = $MR_{G}^{2}$ the guiding center moment of inertia of the ion with respect to the $z$ axis. The cyclotron \textit{spin} part $S_{z} $ is defined as $\left( 2M/q\right) \mu $ with $\mu=m\omega _{c}^{2}\rho _{L}^{2}/2B$ the Larmor magnetic moment. One thus recovers the classical gyromagnetic factor $q/2M$.

Coupling between wave and particle angular momentum components introduced above can be either adiabatic or resonant. At the fluid level, linear adiabatic coupling is described by the Hermitian part of the dielectric tensor, while linear resonant coupling is described by the antihermitian part of the dielectric tensor and the quasilinear equation. Starting with adiabatic coupling, coupling between wave SAM and particles SAM leads to the classical Faraday rotation~\citep{Chen1984,Rax2005}. Adiabatic coupling between wave SAM and particles OAM leads to the mechanical Faraday - or polarisation drag~\citep{Jones1976} - effect whose properties in plasmas have recently been examined~\citep{Gueroult2019a,Gueroult2020}. Lastly, adiabatic coupling between wave OAM and particles OAM leads to the Faraday-Fresnel rotation and splitting recently uncovered for
Trievelpiece-Gould and helicon modes~\citep{Rax2021}. Moving on to resonant coupling, coupling between wave SAM and particles SAM is routinely used for electron
and ion cyclotron resonance heating (ECRH/ICRH) in tokamaks~\citep{Rax2011}, and has also be proposed for mass separation or particle acceleration~\citep{Loeb1986,Pendergast1988,Rax2007,Rax2010}. Meanwhile, as already mentioned in the introduction, resonant coupling between wave OAM and particles OAM has been proposed to control rotation in magnetic mirrors~\citep{Fetterman2008,Fetterman2010} and tokamaks~\citep{Rax2017,Ochs2017b}. 

In this study we will build on and extend these results by deriving the quasilinear kinetic equation which will allow us to identify the exact resonance condition, and from there to uncover couplings between between waves and particles SAM and OAM, in a cylindrical rotating
magnetized plasma. This new resonance condition completes the already identified set of resonant coupling in plasmas : (\textit{i}) Landau in unmagnetized plasmas, (\textit{ii}) cyclotron in magnetized plasmas, and (\textit{iii}) Compton in laser driven plasmas~\citep{Rax1992}. Because our motivation is primarily in rotating mirrors, \textit{straight} tokamaks, and mass filters where the resonant population is the ion population, we will consider a nonrelativistic framework. Under this assumption, we will show that finite Larmor radius effects are responsible for a mixing of OAM and SAM couplings, underlining that rotating magnetized plasmas feature a more complex angular momentum dynamics than unmagnetized plasmas or ordinary neutral matter.

\section{Hamiltonian description of a rotating plasma}
\label{Sec:3}

In this section we lay out the Hamiltonian description of an unperturbed rotation driven by a DC radial electric field in an axially magnetized plasma column. The axial magnetic field is assumed to be produced by a set of coils carrying azimuthal DC currents at the edge of the plasma column. The radial electric field may be generated through DC polarized concentric electrodes at ends of the plasma, provided that the criterion for electric field penetration is fulfilled~\citep{Gueroult2019b,Poulos2019,Liziakin2020,Liziakin2021,Trotabas2022}. Alternatively, in a nonneutral plasma~\citep{Davidson2001}, the electric field is simply the space charge field and there is no need for concentric electrodes. We focus on the ion population in a quasineutral plasma but results can be easily extended to the electron population in quasineutral and nonneutral plasmas.

\subsection{Brillouin modes}

\begin{figure}
\begin{center}
\includegraphics[width = 8cm]{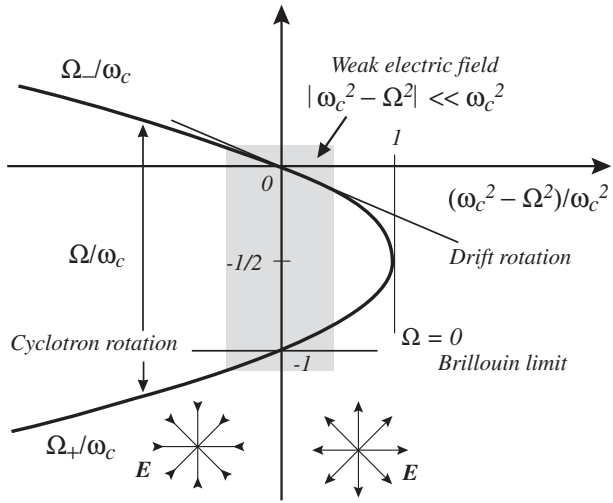}
\caption{The slow and fast angular velocity as a function of the electric field. A clear separation between
guiding center and Larmor radius is relevant for weak electric field in the grey zone.}
\label{Fig:Fig2}
\end{center}
\end{figure}

An ion of mass $M$ and charge $q>0$, interacts with a static radial linear electric field $\mathbf{E}$ and an axial uniform magnetic field $\mathbf{B}$ as shown in figure~\ref{Fig:Fig1} and defined by
\begin{gather}
\frac{q}{M}\mathbf{E} =\left( \frac{\omega _{c}^{2}-\Omega ^{2}}{4}\right)
r~\mathbf{e}_{r},  \label{chan1} \\
\frac{q}{M}\mathbf{B} =\omega _{c}\mathbf{e}_{z}.  \label{chan2}
\end{gather}
Ion orbits are then a combination of the slow and fast Brillouin rotations~\citep{Davidson1969,Davidson2001}. The fast $\Omega _{+}$ and slow $\Omega _{-}$ angular velocities associated with these fast and slow rotations are given by 
\begin{equation}
\Omega _{\pm }  = -\frac{\omega _{c}}{2}\mp \sqrt{\frac{\omega _{c}^{2}}{4}-\frac{qE}{rM}}  \underset{|\frac{E}{rB}| \ll \omega _{c}}{\approx} -\frac{\omega _{c}~}{2}\mp \frac{\omega _{c}}{2}\pm \frac{E}{rB},  \label{res3}
\end{equation}
where $E$ is the DC radial electric field at a given radius $r$\ and $4qE<Mr\omega _{c}^{2}$ is the classical \textit{Brillouin condition}~\citep{Davidson2001}. These two solutions are plotted as a function of the normalized electric field in figure~\ref{Fig:Fig2}.  In the weak electric field regime $\left| E/(rB)\right| \ll \omega _{c}$ highlighted in gray in figure~\ref{Fig:Fig2}, the angular velocity of the guiding center around the $z$ axis reduces to the classical $E\times B$ drift while the angular velocity of the cyclotron motion around the guiding center reduces to the usual cyclotron motion. This corresponds to the asymptotic limit on the right hand side of Eq.~(\ref{res3}). 

From Eq.~(\ref{chan1}) the Brillouin limit leads to the simple requirement $\Omega ^{2}>0$ with
\begin{equation}
\Omega =\sqrt{\frac{q^{2}B^{2}}{M^{2}}-4\frac{qE}{Mr}}.
\label{omeg3}
\end{equation}
Note that $\Omega $, the gyrofrequency $\omega _{c}$ and the wave frequency $\omega $ are all assumed positive throughout this study. Another way to see the condition $\Omega>0$ is to realize that for a given field configuration Eqs.~(\ref{chan1}, \ref{chan2}) the cutoff mass $M^{*}$ between radially unconfined and radially confined ions is the solution of $\Omega ^{2}\left( M^{*}\right) =0$. Ions such that $M<M^{*}$ remain confined around the axis of the configuration. On the other hand ions such that $M>M^{*}$ are expelled radially at an exponential rate. The assumption in this study of $M<M^{*}$ or $\Omega>0$ is thus a requirement to study radially bounded trochoidal orbits. 

With the definition of $\Omega $ in Eq. (\ref{omeg3}) the usual slow and fast Brillouin modes given by Eq. (\ref{res3}) rewrite
\begin{equation}
\Omega _{\pm }=-\frac{\omega _{c}\pm \Omega }{2}.  \label{omeg2}
\end{equation}
One verifies that $\Omega _{+}+\Omega _{-}$ = $-\omega _{c}$ and $\Omega _{+}-\Omega
_{-}$ = $-\Omega $. Note also that $\Omega _{+}<0$ and $\mathbf{\nabla }\cdot 
\mathbf{E}=2\Omega _{+}\Omega _{-}$. The uniform charge density $2\varepsilon _{0}M\Omega _{+}\Omega _{-}/q$ is the small deviation from quasineutrality responsible for the
radial electric field. 

\subsection{Hamiltonian description}

Consider now a system of units such that $q=1$ and $M=1$. In this simple system of units, the electric field and the magnetic field given in Eq.~(\ref{chan1}) and Eq. (\ref{chan2}) derive respectively from the scalar potential 
\begin{equation}
\Phi =\frac{\Omega ^{2}-\omega _{c}^{2}}{8}\left( x^{2}+y^{2}\right) \label{chan4}
\end{equation}
and the vector potential
\begin{equation}
\mathbf{A} =\frac{\omega _{c}}{2}\left( x\mathbf{e}_{y}-y\mathbf{e}_{x}\right).  \label{chan3}
\end{equation}
The unperturbed Hamiltonian $H_{0}$ is classically the sum of the kinetic energy $\mathbf{v}^{2}/2$ plus the potential energy $\Phi \left( \mathbf{r}\right) $, that is
\begin{equation}
H_{0}\left( \mathbf{p},\mathbf{r}\right) =\frac{1}{2}\mathbf{v}^{2}+\Phi =\frac{1}{2}\left[ \mathbf{p-A}\left( \mathbf{r}\right) \right] ^{2}+\Phi\left( \mathbf{r}\right),  \label{h1}
\end{equation}
where $\mathbf{v}$ is the velocity and $\mathbf{p}=p_{x}\mathbf{e}_{x}+p_{y}%
\mathbf{e}_{y}+p_{z}\mathbf{e}_{z}$ is the canonical momentum conjugated to
the position $\mathbf{r}=x\mathbf{e}_{x}+y\mathbf{e}_{y}+z\mathbf{e}_{z}$ of
the ion. In Cartesian coordinates Eq.~(\ref{h1}) rewrites 
\begin{equation}
H_{0}=\frac{1}{2}\left( p_{x}^{2}+p_{y}^{2}\right) +\frac{\omega _{c}}{2}\left( yp_{x}-xp_{y}\right) +\frac{\Omega ^{2}}{8}\left( x^{2}+y^{2}\right) +\frac{p_{z}^{2}}{2}.  \label{h2}
\end{equation}
This is a quadratic form of the Cartesian momentum and positions variables,
so that $H_{0}$ is integrable~\citep{Rax2021a}. 

Let us now introduce the canonical change of variables from the old Cartesian momentum $\left( p_{x},p_{y},p_{z}=P\right) $ and positions $\left(x,y,z\right) $ to the new actions $\left( J,D,P\right) $ and angles $\left( \varphi ,\theta ,z\right) $ variables defined by
\begin{gather}
x =\sqrt{\frac{2D}{\Omega }}\cos \theta -\sqrt{\frac{2J}{\Omega }}\cos
\varphi, \quad y=\sqrt{\frac{2D}{\Omega }}\sin \theta +\sqrt{\frac{2J}{\Omega }}\sin \varphi,  \label{px1} \\
p_{x} =-\sqrt{\frac{\Omega D}{2}}\sin \theta +\sqrt{\frac{\Omega J}{2}}\sin \varphi, \quad p_{y}=\sqrt{\frac{\Omega D}{2}}\cos \theta +\sqrt{\frac{\Omega J}{2}}\cos \varphi,  \label{px2}
\end{gather}
with $J\geq 0$, $D\geq 0$, $\varphi\in\left[ 0,2\pi \right]$ and $\theta\in\left[ 0,2\pi \right] $. By plugging
Eqs.~(\ref{px1}, \ref{px2}) into Eq.~(\ref{h2}) one simply gets
\begin{equation}
H_{0}=-\Omega _{+}J+\Omega _{-}D+\frac{1}{2}P^{2}.  \label{h3}
\end{equation}
This result is independent of the angles $\left( \varphi, \theta, z\right) $ as expected. Note also from Eq.~(\ref{omeg2}) that the cyclotron (kinetic) part of the energy $-\Omega _{+}J$ is always positive, but that the drift (potential) part $\Omega _{-}D$ can be either positive or negative. The particle velocity perpendicular to the magnetic field $\mathbf{v}$ defined as $p_{x}\mathbf{e}_{x}+p_{y}\mathbf{e}_{y}-\mathbf{A}$ and the polar radius $r$ defined as $\sqrt{x^{2}+y^{2}}$ are then obtained from a simple substitution of Eqs.~(\ref{px1}, \ref{px2}) in the Cartesian
definitions, leading to
\begin{multline}
\mathbf{v} =\left( -\Omega _{+}\sqrt{\frac{2J}{\Omega }}\sin \varphi
-\Omega _{-}\sqrt{\frac{2D}{\Omega }}\sin \theta \right) \mathbf{e}_{x}\\
+\left( -\Omega _{+}\sqrt{\frac{2J}{\Omega }}\cos \varphi +\Omega _{-}\sqrt{\frac{2D}{\Omega }}\cos \theta \right) \mathbf{e}_{y}  \label{vvv}
\end{multline}
and
\begin{equation}
r^{2}=x^{2}+y^{2}=2\frac{J+D}{\Omega }-4\frac{\sqrt{JD}}{\Omega }\cos
\left( \theta +\varphi \right).  \label{tri}
\end{equation}

\begin{figure}
\begin{center}
\includegraphics[width = 10cm]{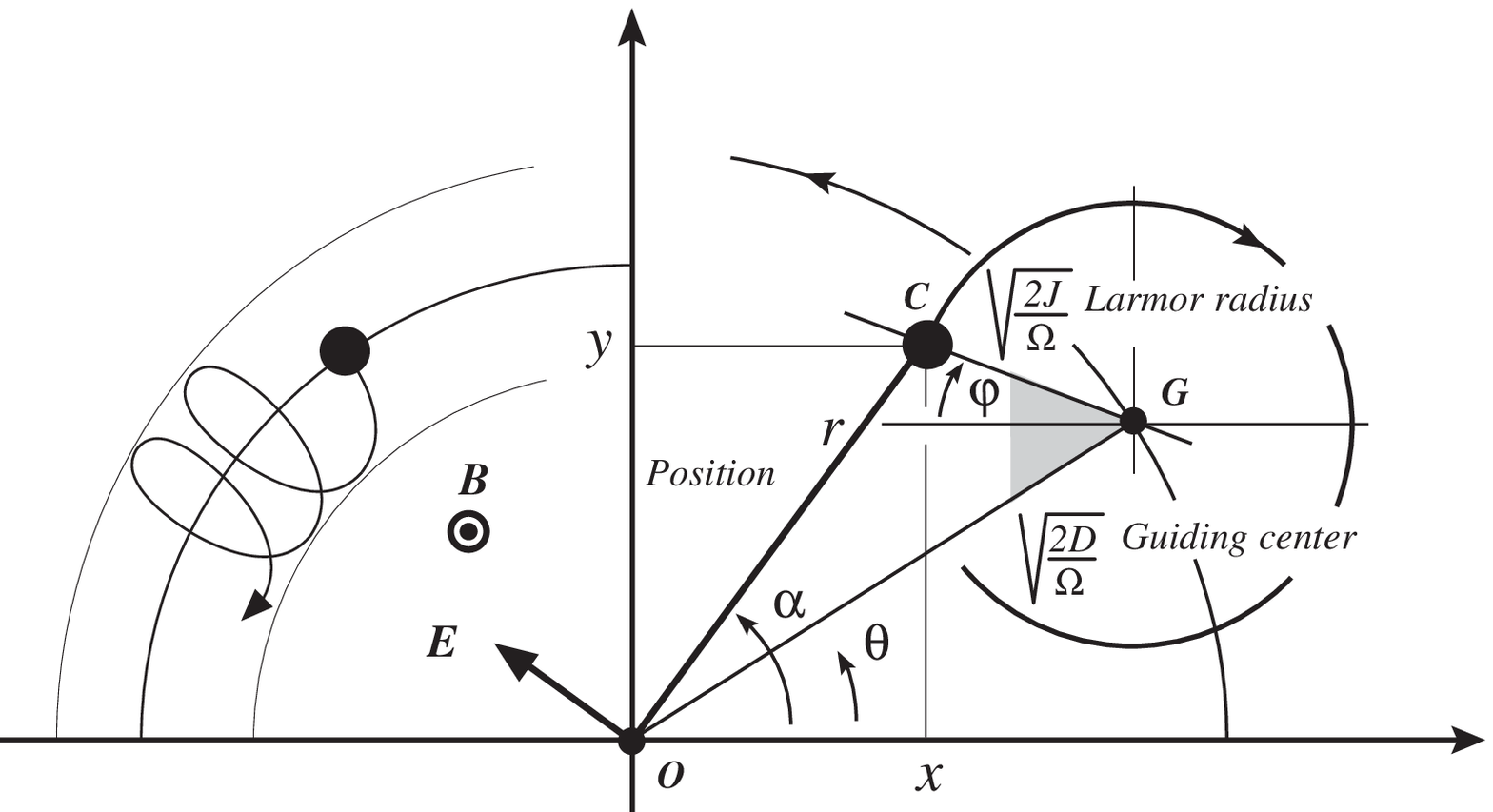}
\caption{Physical meaning of the angle $(\varphi,\theta)$ and actions $(J < D)$ variables in real $(x, y)$ space.}
\label{Fig:Fig3}
\end{center}
\end{figure}

Having identified a set of canonical angles $\left( \varphi ,\theta, z\right) $ and actions $\left( J,D,P\right) $ variables describing the ion interaction with the electric and magnetic field given in Eqs.~(\ref{chan4},~\ref{chan3}), we can now try to shed light onto the physical meaning of the new
variables. Starting with $z$ and $P$, they are  respectively the usual Cartesian coordinate and momentum $P=Mv_{z}$, and their physical interpretation is thus straightforward. The meaning of $\left( J,\varphi \right) $ and $\left(D,\theta \right) $ is on the other hand less obvious. To help our interpretation, figure~\ref{Fig:Fig3} shows the ion motion in the $\left(x,y\right) $ plane when $D>J$. The instantaneous position of the ion is $%
\mathbf{r}=r\mathbf{e}_{r}=\mathbf{OC}$ and it can be viewed as the sum of a \textit{rotating Larmor radius} $\mathbf{GC}$ plus a \textit{rotating guiding center} $\mathbf{OG}$. From figure~\ref{Fig:Fig3},  $\theta$ is the anticlockwise angle between $\mathbf{e}_{x}$ and $\mathbf{OG}$, and $\varphi$ is the clockwise angle between $-\mathbf{e}_{x}$ and $\mathbf{GC}$. We then find from Eq.~(\ref{px1}) that the
guiding center $\left| \mathbf{OG}\right| $ = $\sqrt{2D/\Omega }$, and that the Larmor radius $\left| \mathbf{GC}\right| $ = $\sqrt{2J/\Omega }$. Eq.~(\ref{tri}) is just the \textit{law of cosines} applied to the $\mathbf{OGC}$ triangle with respect to the grey angle in figure~\ref{Fig:Fig3}.

\begin{figure}
\begin{center}
\includegraphics[width= 10cm]{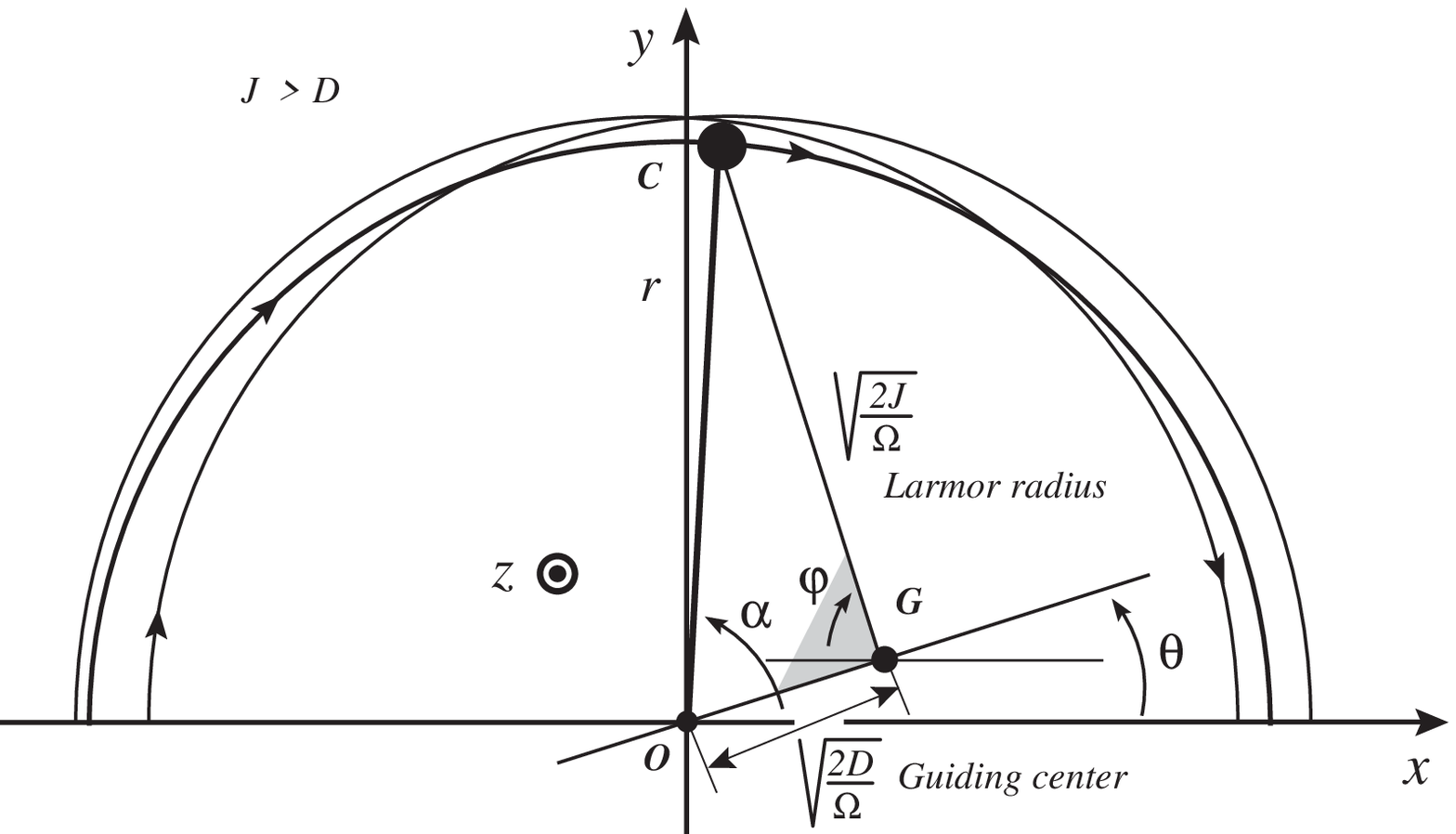}
\caption{Physical meaning of the angle $(\varphi,\theta)$ and actions $(J > D)$ variables in real $(x, y)$ space.}
\label{Fig:Fig4}
\end{center}
\end{figure}

The geometrical interpretation proposed above for $\left( J,\varphi \right) $ and $\left(D,\theta \right)$ based on figure~\ref{Fig:Fig3} assumed $D>J$. If one now considers $J>D$, the canonical description Eqs.~(\ref{px1}, \ref{px2}, \ref{h3}) is still valid, but the picture of the orbit is to be replaced by the one shown in figure~\ref{Fig:Fig4}. As we will show in the next section these two regimes $J\lessgtr D$ can be discriminated based on the sign of the particle \textit{canonical angular momentum}. In effect most of the physical interpretations made in this study will be argued with the ordering $D>J$ in mind as it is the most intuitive, but one should keep in mind that all the relations are valid in both cases $J\lessgtr D$.

Finally, one verifies that Hamilton's equations
\begin{equation}
\frac{d\theta }{dt}=\frac{\partial H_{0}}{\partial D}=\Omega _{-}, \quad%
\frac{d\varphi }{dt}=\frac{\partial H_{0}}{\partial J}=-\Omega _{+},
\end{equation}
lead to the expected classical Brillouin results~\citep{Davidson2001}. The minus sign for the fast (cyclotron) rotation is simply due to the choice of a clockwise angle for $\varphi $ (the counterclockwise choice for $\theta $). It must be stressed here though that the \textit{Larmor radius} angle $\varphi $ does not rotate at the cyclotron frequency $-\Omega _{+}\neq\omega_{c}$. Similarly the $\theta $ angle of the \textit{guiding center} does not rotate with the $E\times B$ velocity $\Omega _{-}\neq -E_{r}/rB$. This is the consequence of inertial effects. The interpretation of the motion as a slow $E\times B$ drift  $\Omega _{-}\approx $ $-E_{r}/rB$ plus a fast cyclotron rotation $\Omega _{+}\approx -\omega _{c}$ is thus only meaningful in the weak electric field limit $\left| E_{r}/B\right| \ll r\omega _{c}$ highlighted in gray in figure~\ref{Fig:Fig2}.

\subsection{Weak field limit}

In order to develop a deeper physical understanding of the weak field regime, which is the one of experimental interest for quasineutral plasmas applications, let us write the ion orbit as a combination of an $E\times B$ slow rotation plus a fast cyclotron rotation. We introduce the guiding center radius $R_{G}$ and the Larmor radius $\rho _{L}$ 
\begin{gather}
x =R_{G}\cos \left( -\frac{E}{rB}t\right) -\rho _{L}\cos (\omega _{c}t),  \label{px3} \\
y =R_{G}\sin \left( -\frac{E}{rB}t\right) +\rho _{L}\sin (\omega _{c}t).  \label{px4}
\end{gather}
Eqs.~(\ref{px3}, \ref{px4}) are simply a rewriting of Eq.(\ref{px1}) in the weak field limits $\left|
E_{r}/rB\right| \ll \omega _{c}$. In this weak field limit $\Omega \approx \omega _{c}$, and the actions $J$ and $D$ can be respectively related to the cyclotron orbit magnetic flux $\Psi _{L}$ = $\pi
\rho _{L}^{2}B$ and the guiding center orbit magnetic flux $\Psi _{G}$ = $\pi R_{G}^{2}B$, with
\begin{gather}
D \approx \frac{\omega _{c}}{2}R_{G}^{2}=\frac{\Psi _{G}}{2\pi },\\
J \approx \frac{\omega _{c}}{2}\rho _{L}^{2}=\frac{\Psi _{L}}{2\pi }.
\end{gather}
Note that these two magnetic fluxes are the first and third adiabatic invariant of Alfven's theory. The usual link between action variables and adiabatic invariants is thus recovered.

Assuming further  $J\ll D$, that is a small Larmor radius, and reintroducing momentarily the ion mass and charge $M$ and $q$ for clarity, one gets in the weak field regime
\begin{gather}
-\Omega _{+}J \approx \frac{M}{2}v_{c}^{2},  \label{tr1} \\
\Omega _{-}D \approx q\Phi \left( R_{G}\right) +\frac{M}{2}\frac{E^{2}}{%
B^{2}}.  \label{tr2}
\end{gather}
Plugging these results into Eq. (\ref{h3}) yields
\begin{equation}
H_{0}\underset{|\frac{E}{rB}| \ll \omega _{c}}{\approx}\frac{M}{2}v_{c}^{2}+\frac{1}{2M}P^{2}+\frac{M}{2}\frac{E^{2}}{B^{2}}+q\Phi.
\label{Eq:Hamiltonian_weak}
\end{equation}
The Hamiltonian thus reduces in this limit to the sum of four terms: the cyclotron kinetic energy $Mv_{c}^{2}/2$, the electrostatic potential energy $q\Phi \left(R_{G}\right) $ of the guiding center, the drift energy $ME^{2}/2B^{2}$ and the parallel kinetic energy .

If one considers now a wave perturbation $\delta H_{0}$ of the Hamiltonian $H_{0}$,  Eq.~(\ref{Eq:Hamiltonian_weak}) shows that in the weak field limit this perturbation is associated with an increase or a decrease of (\textit{i}) the kinetic cyclotron energy $Mv_{c}\delta v_{c}$, (\textit{ii}) the potential energy $q\delta \Phi $ and (\textit{iii}) the axial kinetic energy $Mv_{z}\delta v_{z}$. The structure
of the unperturbed Hamiltonian Eq.~(\ref{h3}) indeed offers in this case the possibility to transfer axial linear momentum, kinetic energy $-\Omega_{+}\delta J\approx Mv_{c}\delta v_{c}$ or/and potential energy $\Omega _{-}\delta D\approx q\delta \Phi $ from rotating waves to rotating particles. When the weak field approximation is no longer valid, inertia effects make this picture more intricate, and the angle-action variables $\left( \varphi ,\theta,z\right) $ and $\left( J,D,P\right) $ then provide the right framework to
understand the dynamics. In order to simplify the algebra, we define $\mathbf{J}$, $\bm{\theta }$ and $\bm{\Omega }$ such that 
\begin{gather}
\mathbf{J} =\left( J,D,P\right)\\
\bm{\theta }=\left( \varphi
,\theta ,z\right), \\
\bm{\Omega } =\frac{d\bm{\theta }}{dt}=\frac{\partial H_{0}}{%
\partial \mathbf{J}}=\left( -\Omega _{+},\Omega _{-},P\right).
\end{gather}
To conclude this section we note that as $H_{0}$ is integrable there exists an infinite set of canonical angle and action variables. The particular choice of \ $\bm{\theta }$ and $\mathbf{J}$ is simply motivated by their
straightforward geometrical interpretation, as illustrated in figure~\ref{Fig:Fig3} and figure~\ref{Fig:Fig4}, and their clear physical meaning in the weak field regime.

\section{Canonical and kinetic angular momentum}
\label{Sec:4}

From Eqs.~(\ref{px1}, \ref{px2}, \ref{vvv}), the $z$ components of the \textit{canonical angular momentum} and of the \textit{kinetic angular momentum} respectively write
\begin{equation}
L_{C}=xp_{y}-yp_{x}=D-J  \label{ccc}
\end{equation}
and
\begin{equation}
L_{K}=xv_{y}-yv_{x}=\frac{2\Omega _{-}}{\Omega }D+\frac{2\Omega _{+}}{\Omega 
}J+\frac{2\omega _{c}}{\Omega }\sqrt{JD}\cos \left( \theta +\varphi \right).
\label{lll}
\end{equation}
As noted in the previous section, the ordering of $J$ and $D$ depends on the sign of $L_{C}$. Specifically, $L_{C}>0$ leads to the orbit topology illustrated in figure~\ref{Fig:Fig3} while $L_{C}<0$ leads to the orbit topology illustrated in figure~\ref{Fig:Fig4}. Note also that $L_{C}$ is independent of time as a consequence of the cylindrical symmetry, but that $L_{K}$ is a function of time since $\theta +\varphi $ = $\left( \Omega _{-}-\Omega_{+}\right) t$.


A physical interpretation of Eq.~(\ref{lll}) can be brought up by considering the moment of inertia of a rotating ion with mass $M=1$. Seeing again the rotating ion as the sum of a rotating guiding center and a rotating Larmor radius, the guiding center moment of inertia is $M_{G}$ = $2D/\Omega $, while the Larmor radius moment of inertia is $M_{L}$ = $2J/\Omega $. The associated angular momentum are $M_{G}d\theta /dt$ = $M_{G}\Omega _{-}$ ($\theta $ is anticlockwise) and $-M_{L}d\varphi /dt$ = $M_{L}\Omega _{+}$ ($\varphi $ is clockwise). One verifies that the sum of these two angular momentum $M_{G}\Omega _{-}$ + $M_{L}\Omega _{+}$ indeed matches $\left\langle L_{K}\right\rangle$ computed from Eq.~(\ref{lll}), where the average $\left\langle {}\right\rangle $ is over the angle $\theta +\varphi $.

Defining the magnetic flux
\begin{equation}
\Psi  = \oint_{\mathcal{C}}\mathbf{A\cdot }d\mathbf{l=}\frac{\omega
_{c}}{2}\oint \left( xdy-ydx\right) = \pi \omega _{c}r^{2}
\label{flux2}
\end{equation}
with $\mathcal{C}$ a contour along a $\Delta \alpha =2\pi $ full turn
of the orbit and $r^{2}$ given by Eq.~(\ref{tri}), Eqs.~(\ref{tri}, \ref{ccc}, \ref{lll}) can be further used to write 
\begin{equation}
L_{C}=L_{K}+\frac{\Psi }{2\pi }.  \label{flux mom}
\end{equation}
We thus see that the criteria $L_{C}\lessgtr 0$, which was identified as determining the type of orbit topology (either that shown in figure~\ref{Fig:Fig3} or that shown in figure~\ref{Fig:Fig4}), can be recasted as $2\pi L_{K}\lessgtr -q\Psi$. This last condition can be interpreted as an ordering between the kinetic energy and the magnetic coupling.

\section{Hamiltonian description of a rotating wave}
\label{Sec:5}

We now consider a wave perturbation associated with a rotating and propagating
potential 
\begin{equation}
\phi \left( \mathbf{r},t\right) =\Real\left[\phi \left( r\right) \exp
j\left( n\alpha +\beta z-\omega t\right) \right],  \label{rf2}
\end{equation}
and a R or L vector potential
\begin{equation}
\mathbf{a}_{\pm }\left( \mathbf{r},t\right) =\Real \left[a\left( r\right)
\exp j\left( n\alpha +\beta z-\omega t\right) \frac{\mathbf{e}_{x}\pm j%
\mathbf{e}_{y}}{\sqrt{2}}\right],  \label{rf1}
\end{equation}
where $n\in \Bbb{Z}$ and $\beta \in \Bbb{R}$. The function $a\left( r\right) 
$ is the real amplitude solution of the radial part of Maxwell-Amp\`{e}re
equation. Maxwell-Faraday equation is fulfilled through $\mathcal{E}%
=-\partial \mathbf{a}_{\pm }/\partial t$ and $\mathcal{B}=\nabla \times 
\mathbf{a}$ where $\mathcal{E}$ and $\mathcal{B}$ are the wave electric and
magnetic fields. The function $\phi \left( r\right) $ is the solution of
Poisson equation. Such solutions of Maxwell-Amp\`{e}re and Poisson equations were recently identified
for the whistler or helicon branch and the Trievelpiece-Gould modes in a
rotating plasma~\citep{Rax2021}. As shown in Appendix A, the wave $\mathbf{a}_{\pm }\left( \mathbf{r},t\right) $ is both an SAM and OAM eigenfunction since 
\begin{equation}
\left( \widehat{L}_{z}+\widehat{S}_{z}\right) \cdot \mathbf{a}_{\pm }=\left(
n\mp 1\right) \mathbf{a}_{\pm },
\end{equation}
and the scalar potential wave $\phi \left( \mathbf{r},t\right) $ is an OAM eigenfunction. 

The perturbed Hamiltonian $H$ describing the interaction of an ion with the DC confining fields Eqs.~(\ref{chan4}, \ref{chan3}) and the rotating RF waves Eqs.~(\ref{rf2}, \ref{rf1}) then writes
\begin{equation}
H=\frac{1}{2}\left( \mathbf{p-A}\right) ^{2}+\Phi -\left( \mathbf{p-A}%
\right) \cdot \mathbf{a}_{\pm }+\frac{\mathbf{a}_{\pm }^{2}}{2}+\phi .
\end{equation}
Here we neglected the second order ponderomotive part of the interaction 
\begin{equation}
\mathbf{a}_{\pm }^{2}/2\ll \mathbf{v}\cdot \mathbf{a}_{\pm }
\end{equation}
but kept the first order dipolar coupling $\mathbf{v}\cdot \mathbf{a}_{\pm }+\phi $
which is responsible for the quasilinear resonant exchange of energy and momentum between waves and particles. This separation between dipolar and ponderomotive perturbations is usual and we will not explore here the interplay between these two couplings~\citep{Ochs2021,Ochs2021a,Ochs2022,Ochs2023} which is associated with the transient build up of the wave. Defining 
\begin{equation}
\mathbf{v}\cdot \mathbf{a}_{\pm }+\phi = v_{x}a_{\pm x}+v_{y}a_{\pm y}+\phi,
\end{equation}
and assuming that $V_{n} \ll H_{0}$, we then write
\begin{equation}
H=H_{0}\left( \mathbf{J}\right) +V_{n}\left( \mathbf{J},\bm{\theta }%
,t\right),  \label{ham54}
\end{equation}
that is that $H$ is decomposed into an unperturbed part $H_{0}$ given in Eq. (\ref{h3}) and the wave perturbation. Using Eqs.~(\ref{vvv}, \ref{rf2}, \ref{rf1}), one gets for the wave pertubation
\begin{multline}
V_{n} =\Real\left[\mp ja\left( r\right) \left[ \Omega _{+}\sqrt{\frac{J}{%
\Omega }}\exp \left( \mp j\varphi \right) -\Omega _{-}\sqrt{\frac{D}{\Omega }%
}\exp \left( \pm j\theta \right) \right]  \exp j\left( n\alpha +\beta z-\omega t\right)\right]\label{ham55} \\
 +\Real\left[\phi \left(
r\right) \exp j\left( n\alpha +\beta z-\omega t\right) \right].
\end{multline}

The last step is to write both $a\left( r\right) $ and $\phi \left(r\right) $ in terms of $\left( J,D\right) $ and $\left( \varphi ,\theta\right) $ to write $V_{n}$ in a form suitable to carry out the quasilinear analysis. This requires finding a convenient basis to express $a\left( r\right) $ and $\phi \left( r\right)$. For quasilinear theory in an homogeneous plasma at rest, this basis is a Fourier set of plane waves associated with translation invariance. Within the framework of Random Phase Approximation (RPA) each Fourier component then acts separately in the quasilinear diffusion operator which is a sum over the square of the amplitude of these Fourier components.

In the case of interest the Fourier-Bessel expansion seems more natural given the cylindrical symmetry of the problem. An added motivation for this choice is that recent studies on the OAM Faraday-Fresnel effect~\citep{Rax2021} have shown that the eigenmodes of the whistler branch in a rotating plasma are of the type described by Eq.~(\ref{rf1}) with in this case $a\left( r\right)$ the ordinary Bessel function $J_{n}\left( kr\right) $, and that the eigenmodes of the Trievelpiece-Gould branch in a rotating plasma are of the type described by
Eq.~(\ref{rf2}) with in this case $\phi \left( r\right)$ proportional to $J_{n}\left(kr\right)$, with $k$ fulfilling in each case an appropriate dispersion relation. Lastly, the Fourier-Bessel expansion theorem states that all the other branches of the plasma waves spectrum in a rotating plasma can similarly be written with waves of the type Eqs.~(\ref{rf2},~\ref{rf1}), and that $a\left( r\right)$ can be represented by $\widetilde{a}\left( k\right) $ as the sum $a\left(r\right) $ = $\int k\widetilde{a}\left( k\right) J_{n}\left( kr\right) dk$ with $\widetilde{a}\left( k\right)=\int ra\left( r\right)J_{n}\left( kr\right) dr$. Thus, without loss of generality, we consider
transverse and longitudinal rotating and propagating cylindrical waves of the type 
\begin{gather}
\mathbf{a}_{\pm }\left( \mathbf{r},t\right) =\Real\left[\frac{\mathcal{E}%
\left( \omega \right) }{j\omega }J_{n}\left( kr\right) \exp j\left( n\alpha
+\beta z-\omega t\right) \frac{\mathbf{e}_{x}\pm j\mathbf{e}_{y}}{\sqrt{2}}\right]%
,  \label{rf22} \\
\phi \left( \mathbf{r},t\right) =\Real\left[\phi \left( \omega \right)
J_{n}\left( kr\right) \exp j\left( n\alpha +\beta z-\omega t\right)\right] .
\label{rf33}
\end{gather}
Here $\mathcal{E}\left( \omega \right) $ and\ $\phi \left( \omega \right)$ are the spectral component of the transverse electric field and potential of a given cylindrical wave packet. The final quasilinear operator, quadratic in $\mathcal{E}$ and $\phi $, will ultimately be summed over the full $\omega $ and $k\left( \omega \right) $ spectra. In this study though we do not specify the $\omega $ spectra and the $k\left( \omega,\beta ,n\right) $ dispersion. To provide a general result we instead analyze the quasilinear dynamics under the influence of a single cylindrical component Eqs.~(\ref{rf22}, \ref{rf33}). The quasilinear effect of a wave packet is simply the sum over the effects of each Fourier-Bessel component as in the plane wave case.

In a homogeneous magnetized plasma at rest the Jacobi-Anger's expansion is used to identify the harmonic cyclotron resonances of a plane wave. For the cylindrical waves described by Eqs.~(\ref{rf22}, \ref{rf33}) we instead consider the triangle $\mathbf{OGC}$ in figure~\ref{Fig:Fig3} and figure~\ref{Fig:Fig4} and apply Graf's addition theorem 
\begin{equation}
J_{n}\left( kr\right) \exp \left( jn\alpha \right) =\sum_{l=-\infty}^{l=+\infty }J_{l+n}\left( k\sqrt{\frac{2D}{\Omega }}\right) J_{l}\left( k\sqrt{\frac{2J}{\Omega }}\right) \exp j\left[ \left( l+n\right) \theta
+l\varphi \right] .
\label{Eq:Graf}
\end{equation}
In the limit of zero Larmor radius the sum on the right hand side reduces to the $l=0$ term only, whereas $l\neq 0$ terms capture finite larmor radius effects. One verifies that the larger $k$ and the larger $J/D$, the more terms are needed to approach the left hand side in Eq.~(\ref{Eq:Graf}). Plugging this result into Eqs.~(\ref{ham54}, \ref{ham55}) the Hamiltonian writes 
\begin{equation}
H=H_{0}\left( \mathbf{J}\right) +\Real\left[\sum_{l=-\infty }^{l=+\infty
}V_{nl\sigma }\left( J,D\right) \exp j\left[ \left( l+\sigma \right) \varphi
+\left( l+n\right) \theta +\beta z-\omega t\right] \right]  \label{ham1213}
\end{equation}
where we introduced $\sigma \in \left[ -1,0,+1\right] $ the SAM index such that the L, R and scalar wave perturbations $V_{nl\sigma }$ are given by 
\begin{align}
V_{n,l,\sigma =+1} &=\mathcal{E}\frac{\Omega _{+}}{\omega }\sqrt{\frac{J}{%
\Omega }}J_{l+n}\left( k\sqrt{\frac{2D}{\Omega }}\right) J_{l}\left( k\sqrt{%
\frac{2J}{\Omega }}\right)   \nonumber \\
&\qquad-\mathcal{E}\frac{\Omega _{-}}{\omega }\sqrt{\frac{D}{\Omega }}%
J_{l+1+n}\left( k\sqrt{\frac{2D}{\Omega }}\right) J_{l+1}\left( k\sqrt{\frac{%
2J}{\Omega }}\right) , \label{ph1}\\
V_{n,l,\sigma =-1} &=\mathcal{E}\frac{\Omega _{-}}{\omega }\sqrt{\frac{D}{%
\Omega }}J_{l-1+n}\left( k\sqrt{\frac{2D}{\Omega }}\right) J_{l-1}\left( k%
\sqrt{\frac{2J}{\Omega }}\right)  \nonumber\\
&\qquad-\mathcal{E}\frac{\Omega _{+}}{\omega }\sqrt{\frac{J}{\Omega }}%
J_{l+n}\left( k\sqrt{\frac{2D}{\Omega }}\right) J_{l}\left( k\sqrt{\frac{2J}{%
\Omega }}\right) ,  \label{ph2} \\
V_{n,l,\sigma =0} &=\phi J_{l+n}\left( k\sqrt{\frac{2D}{\Omega }}\right)
J_{l}\left( k\sqrt{\frac{2J}{\Omega }}\right) .  \label{ph3}
\end{align}

In order to simplify the notation we define the vector
\begin{equation}
\mathbf{N}=\left[ \left( l+\sigma \right) ,\left( l+n\right) ,\beta \right] 
\end{equation}
and use it as an index to specify $n$, $l$ and $\sigma$. The Hamiltonian $H$ in Eq. (\ref{ham54}) then writes in compact form
\begin{equation}
H\left( \mathbf{J},\bm{\theta },t\right) =H_{0}\left( \mathbf{J}\right)
+\sum_{\mathbf{N}}V_{\mathbf{N}}\left( \mathbf{J}\right) \exp j\left( 
\mathbf{N}\cdot \bm{\theta}-\omega t\right) ,  \label{ham89}
\end{equation}
where we dropped the $\Real$ mention for readability and $\sum_{\mathbf{N}}=
\sum_{l=-\infty }^{l=+\infty }$. This implies that both the OAM azimuthal number $n$ of the cylindrical waves Eqs.~(\ref{rf22}, \ref{rf33}) and the SAM number $\sigma $ remain fixed as we study separately L,
R and potential waves.

For a purely rotating wave, characterized by $n$, $\sigma $ and $\beta =0$, the structure of the relation Eq.~(\ref{ham89}) reveals a number of selection rules between the wave induced small increments of actions and energy. These selection rules provide also the branching ratio between the exchange of the actions. To see this consider Hamilton's equations
\begin{gather}
\frac{dH}{dt} =\frac{\partial H}{\partial t}=-j\omega \sum_{\mathbf{N}}V_{%
\mathbf{N}}\left( \mathbf{J}\right) \exp j\left( \mathbf{N}\cdot \mathbf{%
\theta }-\omega t\right) \\
\frac{dL_{C}}{dt} =\frac{\partial H}{\partial \varphi }-\frac{\partial H}{%
\partial \theta }=j\left( \sigma -n\right) \sum_{\mathbf{N}}V_{\mathbf{N}%
}\left( \mathbf{J}\right) \exp j\left( \mathbf{N}\cdot \bm{\theta}%
-\omega t\right)
\end{gather}
A wave induced small variation of the ion energy $\delta H$ thus implies a small variation of the angular momentum $\delta L_{C}$ through
\begin{equation}
\frac{\delta L_{C}}{\delta H}=\frac{n-\sigma }{\omega }.  \label{sru1}
\end{equation}
Using $\delta H=\Omega _{-}\delta D-\Omega _{+}\delta J$ we can express the
branching ration between the cyclotron kinetic energy channel and the
potential energy channel as 
\begin{gather}
\frac{\delta J}{\delta H} =\frac{l+\sigma }{\omega },  \label{brs1}
\\
\frac{\delta D}{\delta H} =\frac{l+n}{\omega }.  \label{brs3}
\end{gather}
These relations provide a first basic tool to optimize phase space engineering. For example one may want to set up a radial current ($\delta D\neq 0$) but avoid ICRH heating ($\delta J=0$). In this case Eqs.~(\ref{brs1}, \ref{brs3}) point to waves such that $l+\sigma=0$ but $l+n\neq 0$.  However the best strategy to optimize power deposition among the various energy channels is to consider the kinetic equation.

\section{Brillouin resonances}
\label{Sec:6}

\subsection{Resonance condition}
\label{Subec:resonance}

The Hamiltonian Eq. (\ref{ham89}) makes it possible to identify the conditions for resonant coupling. For this we simply substitute the unperturbed motion $\bm{\theta }= \bm{\Omega }t + \bm{\theta }_{0}$ into the oscillating phase $j\left( \mathbf{N}\cdot \bm{\theta }-\omega t\right) $ of each $V_{\mathbf{N}}\exp j\left( \mathbf{N}\cdot \bm{\theta }-\omega t\right)$ perturbation. We then obtain the phase factor $j\left( 
\mathbf{N}\cdot \bm{\Omega}-\omega \right) t$ + $j\bm{\theta }_{0}$ which
stops to rotate and becomes stationary when the resonance
condition 
\begin{equation}
\omega = \mathbf{N}\cdot \bm{\Omega} =-\left( l+\sigma \right) \Omega _{+}+\left(
l+n\right) \Omega _{-}+\beta P.  \label{res 65}
\end{equation}
is fulfilled. This resonance condition replaces the classical Landau-cyclotron condition
\begin{equation}
\omega -\beta v_{z}=m\omega _{c}  \label{res1}
\end{equation}
with $m \in \mathbb{Z}$ and where $v_{z}$ is the velocity along the magnetic field. Because this is the slow and fast Brillouin modes that are involved in Eq.~(\ref{res 65}) rather than the cyclotron frequency we call Eq.~(\ref{res 65}) the \textit{Brillouin} resonance condition. One verifies as expected that Eq.~(\ref{res 65}) reduces to Eq.~(\ref{res1}) when the static electric field Eq. (\ref{chan1}) cancels. Rewriting Eq.~(\ref{res 65}) as 
\begin{equation}
\omega -\beta P-(n+\sigma)\Omega_{-} = l(\omega_{c}+2\Omega_{-}) +\sigma \omega_{c},
\end{equation}
the left hand side is simply the Doppler shifted wave frequency considering both the axial translation and the azimuthal rotation~\citep{Garetz1981,Courtial1998}. The first term on the right hand side $l(\omega_{c}+2\Omega_{-})$ can then be interpreted as normal ($l>0$) or anomalous ($l<0$) Doppler effect modified by inertial effects. Indeed $\omega_{c}+2\Omega_{-}$ is the gyro-frequency corrected by the Coriolis force for an ion rotating at the angular frequency $\Omega_{-}$. The classical yet subtle picture of normal/anomalous Doppler effect can be extended to the cases where the helical motion of the guiding center ($P,\Omega _{-}$) is slower or faster than the axial ($\omega /\beta $) and azimuthal ($\omega /n$) phase velocities~\citep{Nezlin1976}.

At resonance $\exp j\left( \mathbf{N}\cdot \bm{\theta }-\omega\right)t= \exp j( \mathbf{N}\cdot \bm{\theta}_{0})$. Some particles gain energy-momentum while others loose energy-momentum, with the sign of the variation determined by the phase factor $\Real\left(\exp j( \mathbf{N}\cdot \bm{\theta}_{0})\right)$. This diffusive behaviour of the actions is described within the framework of Random Phase Approximation (RPA) where we average over $\exp j( \mathbf{N}\cdot \bm{\theta}_{0})$ the square of the action variations to construct the quasilinear diffusion equation. The construction of this kinetic description leading to the quasilinear equation can done either through a Lagrangian or an Eulerian point of view in phase space~\citep{Rax2021a}. Here we will use the latter, as reviewed in Appendix~\ref{Sec:AppendixB}.


Note finally that the resonance condition Eq.~(\ref{res 65}) can be recovered from a simple photon picture. For this we recall that a photon associated with a wave described by Eq.~(\ref{oam1}) carries an energy $\hbar \omega $ and a linear momentum along the $z$ axis $\hbar\beta $. When this photon is absorbed by a rotating ion the variation of the particle energy $H_{0}$ and linear momentum $P$ are given by 
\begin{equation}
\delta H_{0}=\hbar \omega ,\quad\delta P=\hbar \beta .
\end{equation}
Besides energy and axial linear momentum, the photon associated with the wave Eq.(\ref{oam1}) also carries an OAM plus SAM angular momentum $\left(n\mp1\right)\hbar$ along the $z$ axis (see Appendix~\ref{Sec:SAMOAM}). When this photon is absorbed by a rotating ion the change of the ion canonical angular momentum $L_{C}$ is 
\begin{equation}
\delta L_{C}=\left( n\mp 1\right) \hbar .  \label{phot}
\end{equation}
Eq.~(\ref{h2}) reveals the harmonic oscillators structure of the Hamiltonian $H_{0}$. We can thus draw an analogy with the Hamiltonian of the Landau levels of a magnetized quantum particle to conclude that, at
the quantum level, the changes of the action $J$ and $D$ can only be an integer multiple of $\hbar$, that is
\begin{equation}
\delta J=n_{J}\hbar ,\quad\delta D=n_{D}\hbar 
\end{equation}
with $\left( n_{J},n_{D}\right) \in \Bbb{Z}^{2}$  a pair of integers. In fact we are considering the quasi-classical limit with large quantum numbers : $n_{J}+1/2\sim n_{J}$ and $n_{D}+1/2\sim n_{D}$. Then, from Eqs.~(\ref{h3}, \ref{ccc}), one finds 
\begin{gather}
\delta L_{C} = \delta D - \delta J,\\
\delta H_{0} = \Omega _{-}\delta D-\Omega _{+}\delta J +\delta P^{2}/2.
\end{gather}
These two relations together with the semiclassical expansion $\delta P^{2}$ = $2\hbar \beta P+O\left( \hbar ^{2}\right)$ finally lead to
\begin{equation}
\omega -\beta P=-n_{J}\Omega _{+}+\left( n_{J}+n\mp 1\right) \Omega
_{-}+O\left( \hbar \right)   \label{res9}
\end{equation}
If $\hbar = 0$, we recognize a relation similar to our resonance condition identified in Eq. (\ref{res 65}) with $n_{J}$ = $l\pm 1$, supporting the simple photon picture.

\subsection{Diffusion paths}

Besides the resonance lines Eq.~(\ref{res 65}), we can also identify from the Hamiltonian the diffusion paths along which resonant energy-momentum exchanges take place in actions space. Restricting the study to a single $V_{\mathbf{N}}\left(\mathbf{J}\right)$ resonant coupling in Eq.~(\ref{ham89}), Hamilton's equations near the resonance Eq. (\ref{res 65}) write
\begin{gather}
\left. \frac{d\mathbf{J}}{dt}\right| _{\mathbf{N\cdot \bm{\Omega} }=\omega } =-j%
\mathbf{N}V_{\mathbf{N}}\exp j\mathbf{N}\cdot \bm{\theta }_{0}, \\
\left. \frac{dH}{dt}\right| _{\mathbf{N\cdot \bm{\Omega} }=\omega } =-j\omega
V_{\mathbf{N}}\exp j\mathbf{N}\cdot \bm{\theta }_{0}.
\end{gather}
Resonant actions variation (wave kicks) $\delta \mathbf{J}$ associated with a resonant energy variation $\delta H$ are thus related by 
\begin{equation}
\omega \delta \mathbf{J}=\mathbf{N}\delta H.  \label{diffpath}
\end{equation}
Practically $\delta H$ can be taken as $\delta H_{0}$ within the two time scales quasilinear framework reviewed in Appendix~\ref{Sec:AppendixB}. In fact, without invoking the ordering between the secular quasilinear evolution and the fast $\omega$ oscillation, we can simply evaluate a variation of $H_{0}$ in Eq.~(\ref{h3}) and take into account Eq. (\ref{diffpath}) to obtain 
\begin{equation}
\delta H_{0}=\bm{\Omega }\cdot \delta \mathbf{J}=\bm{\Omega }\cdot 
\mathbf{N}\frac{\delta H}{\omega }=\delta H
\end{equation}
since $\omega =\mathbf{N}\cdot \bm{\Omega}$. Thus the resonant wave kicks $\delta\mathbf{J}$ associated with a resonant energy exchange $\delta H_{0}$ between the rotating wave and the rotating particle are given by 
\begin{equation}
\left[ \delta J,\delta D,\delta P\right] =\left[ \left( l+\sigma \right)
,\left( l+n\right) ,\beta \right] \frac{\delta H_{0}}{\omega }
\label{diffpath33}
\end{equation}
for a given $V_{\mathbf{N}}$ coupling. In the weak field limit this reduces to
\begin{equation}
\left[ \omega_{c}\rho_{L}\delta\rho_{L},\omega_{c}R_{g}\delta R_{g},\delta v_{z}\right] =\left[ \left( l+\sigma \right)
,\left( l+n\right) ,\beta \right] \frac{\delta H_{0}}{M\omega }
\end{equation}
where we have temporarily reintroduced the ion mass $M$ and the cyclotron frequency $\omega_{c}$. 

\begin{figure}
\begin{center}
\includegraphics[height = 7cm]{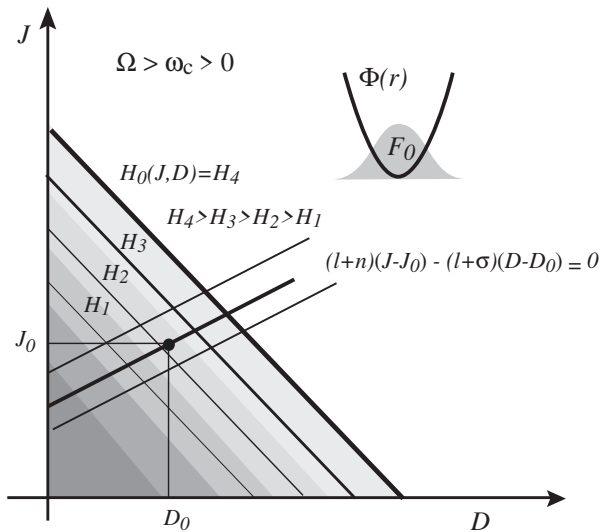}
\caption{Isoenergy levels $H_0$ and diffusion paths in $(J, D)$ action space.}
\label{Fig:Fig5}
\end{center}
\end{figure}

The relation Eq. (\ref{diffpath}) implies that there exists a linear combination of the actions which is invariant
under the time evolution prescribed by the wave coupling $V_{\mathbf{N}}\exp j\left( \mathbf{N}\cdot \bm{\theta }-\omega t\right) $, namely
\begin{equation}
\delta \left( \mathbf{N}\times \mathbf{J}\right) =0.  \label{inv1}
\end{equation}
Focusing on Brillouin resonances rather than on Landau resonances,  we restrict the analysis to the case $\beta P=0$. For a single $V_{\mathbf{N}}$ Eqs.~(\ref{diffpath}, \ref{inv1}) then identify a diffusion path in action space $(J,D)$. Specifically, the path passing through a resonant point $\left(J_{0},D_{0}\right)$ writes
\begin{equation}
\left( l+n\right) \left( J-J_{0}\right) =\left( l+\sigma \right) \left(
D-D_{0}\right).   \label{diffpath2}
\end{equation}
This is illustrated in figure~\ref{Fig:Fig5}. The diffusion paths Eq.~(\ref{diffpath2}) are invariant under the dynamics driven by a single $V_{\mathbf{N}}$ coupling. Quasilinear diffusion in action space takes place along these diffusion paths provided that the resonant condition Eq. (\ref{res 65}) is fulfilled and that $\left|V_{nl\sigma }\left( J_{0},D_{0}\right) \right| $ is not too small. Figure~\ref{Fig:Fig5} also represents the isoenergy lines, \emph{i.~e.} points in $\left(J,D\right)$ space where $H_{0}$ is constant. For $\Omega >\omega _{c}>0$ the electric field is confining and we can consider a canonical equilibrium distribution function
\begin{equation}
F_{0}\left( \mathbf{J}\right) =\frac{\Omega ^{2}-\omega _{c}^{2}}{4\sqrt{%
2\pi }\left( k_{B}T\right) ^{\frac{5}{2}}}\exp \left(-\frac{H_{0}\left( \mathbf{J}%
\right) }{k_{B}T} \right) \label{max}
\end{equation}
where $T$ is the temperature and we have taken the normalization $\int d\mathbf{J}F_{0}=1$. The corresponding density levels in $(J,D)$ space are color coded in gray in figure~\ref{Fig:Fig5}. For $0<\Omega \leq\omega _{c}$ the canonical distribution function Eq.~(\ref{max}) cannot be normalized as the potential $\phi(r)$ is not confining but instead either flat ($\Omega=\omega_{c}$) or hill shaped ($\Omega<\omega_{c}$).

\subsection{Small Larmor radius limit}

When the Larmor radius $\sqrt{2J/\Omega }$ is small compared to the radial wavelength $2\pi/k$, that is 
\begin{equation}
k\sqrt{\frac{J}{2\Omega} }<1,
\end{equation}
the coupling coefficients $V_{n,l,\sigma }$ in Eqs.~(\ref{ph1}, \ref{ph2}, \ref{ph3}) can be simplified by using the small parameter expansion 
\begin{equation}
J_{l}\left( k\sqrt{\frac{J}{2\Omega} }\right) \sim \frac{\left( k\sqrt{\frac{J}{2\Omega} }\right) ^{l}}{l!}.
\end{equation}
We see that in this limit the $l=0$ term is the most effective since it is associated with the
largest $V_{n,l,\sigma }$. The resonance condition Eq. (\ref{res 65}) for $l=0$ then gives $\omega -\beta P$ = $-\sigma\Omega _{+}+n\Omega _{-}$ and the fundamental $l=0$ Doppler shifted cyclotron resonance term $\sigma \Omega _{+}$ is only due to the SAM content of the wave. 

If one assumes further $\sigma =0$, as in the scalar case Eq. (\ref{rf33}), and $\beta P=0$, cyclotron and Landau terms are avoided and one gets get a pure OAM harmonic Brillouin-Landau resonance  
\begin{equation}
\omega =n\Omega _{-}  \label{ressc}
\end{equation}
between the drift rotation $\Omega_{-}$ and the wave OAM. This is the optimal choice for the sustainment of a plasma column rotation. The physical interpretation of Eq. (\ref{ressc}) is simple : the wave angular velocity $d\alpha /dt$ = $\omega /n$ is equal to the particle guiding center angular velocity $-d\theta /dt$ = $\Omega _{-}$. 

\subsection{Special examples}

Let us illustrate the physics behind Brillouin coupling through three selected examples. Consider first a potential wave $\sigma =0$, that is Eq.~(\ref{rf33}), with $\beta =0$ and the small Larmor radius approximation such that $l=0$. In these conditions $\mathbf{N}$ = $\left[ 0,n,0\right] $ and Eq.~(\ref{diffpath33}) writes 
\begin{equation}
\frac{\delta J}{\delta H_{0}}=0,\quad\frac{\delta D}{\delta H_{0}}=\frac{1%
}{\Omega _{-}}  \label{eff12}
\end{equation}
since the Brillouin resonance Eq.~(\ref{res 65}) implies $\omega $ = $n\Omega _{-}$. This result can be interpreted as follows. For absorption $\delta H_{0}>0$, the wave energy is transferred to potential
energy through the wave induced radial dynamics of the guiding center in the electrostatic potential $\Phi\left(r\right) $. For emission $\delta H_{0}<0$, the potential
energy of the particles is passed on to the wave energy. This type of instability is used for microwave generation in magnetrons. We note also that the simultaneous cooling ($\delta H_{0}<0$) and ash removal ($\delta D>0$) of alpha particles in a rotating tokamak, through free energy extraction~\citep{Fisch1992,Fisch1993,Fisch1994,Fisch1995,Herrmann1997}, is optimal in these wave conditions. The ratio of energy extraction to radial expulsion is then adjusted through $\Omega _{-}$.

Consider now a vectorial wave $\sigma =\pm 1$, that is Eq. (\ref{rf22}), with $\beta =0$, no OAM ($n=0$)
and the small Larmor radius approximation such that $l=0$. In these conditions $\mathbf{N} = \left[ \sigma ,0,0\right] $ and Eq.~(\ref{diffpath33}%
) writes
\begin{equation}
\frac{\delta D}{\delta H_{0}}=0,\quad\frac{\delta J}{\delta H_{0}}=-\frac{%
1}{\Omega _{+}}  \label{LCLC1}
\end{equation}
since the Brillouin resonance Eq. (\ref{res 65}) implies $\omega = -\sigma\Omega _{+}$ (remember that $\Omega _{+}<0$). In this case the wave energy is simply passed into cyclotron energy $-\Omega
_{+}\delta J$. This is the case of pure ICRH modified by inertial effects.

Consider finally a potential wave $\sigma =0$, that is Eq. (\ref{rf33}), with $\beta =0$ and no OAM ($n=0$), but with finite Larmor radius effects $l\neq 0$. In these conditions $\mathbf{N}$ = $\left[ l,l,0\right] $ and Eq.~(\ref{diffpath}) writes
\begin{equation}
\frac{\delta D}{\delta H_{0}}=\frac{1}{\Omega },\quad\frac{\delta J}{%
\delta H_{0}}=\frac{1}{\Omega }
\end{equation}
since the Brillouin resonance Eq. (\ref{res 65}) implies $\omega $ = $-l\Omega_{+}$ $+$ $l\Omega _{-}$ = $l\Omega $. This result can be interpreted as follows. For absorption $\delta H_{0}>0$ the wave energy is transferred to both kinetic energy $-\Omega _{+}\delta J$ and potential energy $\Omega_{-}\delta D$ as shown by Eq. (\ref{h3}). The partitioning between
these two energy channels must however fulfils conservation of canonical angular momentum $\delta L_{C}=0$, which from Eq. (\ref{ccc}) implies $\delta D= \delta J$. Conservation of $L_{C}$ is indeed the consequence of the invariance through rotation of the Hamiltonian $H_{0}+V$, since here the wave carries neither OAM ($n=0$) nor
SAM ($\sigma =0$).


\section{Quasilinear theory in a rotating magnetized plasma}
\label{Sec:7}

When a wave propagates in a plasma two times scales are associated with the quasilinear ordering. One is the fast linear response that is described at the kinetic level by Vlasov's equation, whose solution gives the
refractive/dispersive part of the dielectric tensor. The other is the slow, angle averaged, quasilinear evolution of the action distribution function $F\left( \mathbf{J},t\right) $ which is described at the kinetic level by the quasilinear equation. The standard quasilinear equation Eq.~(\ref{ql568}) is derived in Appendix~\ref{Sec:AppendixB}. 

With the Hamiltonian Eq.~(\ref{ham89}), the evolution of the action distribution function $F\left( \mathbf{J},t\right) $ in a magnetized rotating plasmas is given by 
\begin{equation}
\frac{\partial F}{\partial t}=\frac{\pi }{2}\sum_{\mathbf{N}}\left( \mathbf{%
N\cdot }\frac{\partial }{\partial \mathbf{J}}\right) \left[ \left| V_{%
\mathbf{N}}\right| ^{2}\delta \left( \mathbf{N}\cdot \bm{\Omega}-\omega \right)
\right] \left( \mathbf{N}\cdot \frac{\partial }{\partial \mathbf{J}}\right)
F\left( \mathbf{J},t\right)   \label{ql}
\end{equation}
where the $V_{\mathbf{N}}$ are those derived in Eqs.~(\ref{ph1}, \ref{ph2}, \ref{ph3}).
The operator $\mathbf{N}\cdot \partial_{\mathbf{J}}$ involved in this slow (with respect to $1/\omega $) diffusion in action space writes
\begin{equation}
\mathbf{N\cdot }\frac{\partial }{\partial \mathbf{J}}=\left( l+\sigma\right) \frac{\partial }{\partial J}+\left( l+n\right) \frac{\partial }{\partial D}+\beta \frac{\partial }{\partial P}.  \label{opq}
\end{equation}
We normalize $F$ by taking $\int d\mathbf{J}F\left( \mathbf{J},t\right) =1$ and consider that $F\left( \mathbf{J}=+\infty ,t\right) =0$ and $F\left(J<0,D<0,P=-\infty ,t\right) =0$. 

Introducing the resonant particles density $\rho $, the power per unit volume $W$ lost or gained by the wave (and
gained or lost by the plasma) is 
\begin{equation}
\frac{W}{\rho} =\frac{d}{dt}\int d\mathbf{J}H_{0}\left( \mathbf{J}\right)
F\left( \mathbf{J},t\right) =\int d\mathbf{J}H_{0}\frac{\partial F}{\partial
t} \label{epr}
\end{equation} 
where the integral is to be taken over $-\infty <J,D,P<+\infty$. Using Eq.~(\ref{ql}) and integrating by part the operator Eq.~(\ref{opq}) gives
\begin{equation}
\frac{W}{\rho} = -\frac{\pi }{2}\int d\mathbf{J}\omega \sum_{\mathbf{N}}\left[ \left| V_{%
\mathbf{N}}\right| ^{2}\delta \left( \mathbf{N}\cdot \bm{\Omega}-\omega \right)
\right] \mathbf{N}\cdot \frac{\partial F}{\partial \mathbf{J}}  \label{epr2}
\end{equation}
We recognize in Eq. (\ref{epr2}) the power balance given by the dissipative part of the collisionless dielectric
tensor. A second integration by part then gives the density of power 
\begin{equation}
W= \frac{\pi }{2}\rho\int d\mathbf{J}F\left( \mathbf{J},t\right) \omega
\sum_{\mathbf{N}}\mathbf{N\cdot }\frac{\partial }{\partial \mathbf{J}}\left[
\left| V_{\mathbf{N}}\right| ^{2}\delta \left( \mathbf{N}\cdot \bm{\Omega}%
-\omega \right) \right].   \label{poww}
\end{equation}
Defining $w_{\mathbf{N}}\left( \mathbf{J}\right)$ through
\begin{equation}
W = \rho \sum_{\mathbf{N}}\int d\mathbf{J}F\left( \mathbf{J}\right) w_{\mathbf{N}}\left( \mathbf{J}\right),
\end{equation}
Eq.~(\ref{poww}) then gives 
\begin{equation}
w_{\mathbf{N}}\left( \mathbf{J}\right) =\frac{\pi }{2}\omega \sum_{\mathbf{N}}\mathbf{N}\cdot \frac{\partial }{\partial \mathbf{J}}\left[ \left| V_{\mathbf{N}}\right| ^{2}\delta \left( \mathbf{N}\cdot \bm{\Omega}-\omega \right)
\right].   \label{einbh}
\end{equation}
This quantity is thus interpreted, for a single component $V_{\mathbf{N}}$, as the power exchanged by a particle at $\mathbf{J}$ with this $V_{\mathbf{N}}$ component of the wave interaction. Eq.~(\ref{diffpath}) can hence be rewritten as 
\begin{equation}
\left. \frac{\left\langle \delta \mathbf{J}\right\rangle }{\delta t}\right|
_{\mathbf{N}}=\frac{\mathbf{N}}{\omega }\frac{\delta H_{0}\left( \mathbf{J}\right) }{\delta t}=\frac{\mathbf{N}}{\omega }w_{\mathbf{N}}\left( \mathbf{J}\right),   \label{einhjy}
\end{equation}
where $\left\langle {}\right\rangle $ is an average over the fast phase of the wave and the $\mathbf{N}$ index restricts the wave kick $\delta \mathbf{J}$ to a single component $V_{\mathbf{N}}$.  Plugging Eq.~(\ref{einbh}) into Eq.~(\ref{einhjy}) and using the tensorial product notation $\mathbf{\otimes }$ yields 
\begin{equation}
\frac{\left\langle \delta \mathbf{J}\right\rangle }{\delta t}=\frac{\pi }{2}\frac{\partial }{\partial \mathbf{J}}\cdot \sum_{\mathbf{N}}\mathbf{N\otimes
N}\left| V_{\mathbf{N}}\right| ^{2}\delta \left( \mathbf{N}\cdot \bm{\Omega}-\omega \right)   \label{einhjy5}
\end{equation}
This relation Eq. (\ref{einhjy5}) is just the usual Einstein's relation~\citep{Fraiman1995} between the wave induced drift coefficient $\left\langle \delta \mathbf{J}\right\rangle/\delta t$ and the wave induced diffusion coefficient
\begin{equation}
\frac{\left\langle \delta \mathbf{J\otimes }\delta \mathbf{J}\right\rangle }{2\delta t}=\frac{\pi }{2}\sum_{\mathbf{N}}\mathbf{N\otimes N}\left[ \left|
V_{\mathbf{N}}\right| ^{2}\delta \left( \mathbf{N}\cdot \bm{\Omega}-\omega
\right) \right] 
\end{equation}
used to write the kinetic equation Eq.~(\ref{ql}) in the classical Fokker-Planck form with a drift and a diffusion coefficient 
\begin{equation}
\frac{\partial F}{\partial t}=-\frac{\partial }{\partial \mathbf{J}}\cdot
\left[ \frac{\left\langle \delta \mathbf{J}\right\rangle }{\delta t}F-\frac{\partial }{\partial \mathbf{J}}\cdot \frac{\left\langle \delta \mathbf{J\otimes }\delta \mathbf{J}\right\rangle }{2\delta t}F\right] 
\end{equation}
rather than in the completely equivalent and more usual quasilinear form 
\begin{equation}
\frac{\partial F}{\partial t}=\frac{\partial }{\partial \mathbf{J}}\cdot
\left[ \frac{\left\langle \delta \mathbf{J\otimes }\delta \mathbf{J}\right\rangle }{2\delta t}\cdot \frac{\partial F}{\partial \mathbf{J}}\right] 
\end{equation}
used here in Eq.~(\ref{ql}). Einstein's relation 
\begin{equation}
\frac{\left\langle \delta \mathbf{J}\right\rangle}{\delta t} = \partial_{\mathbf{J}}\cdot\frac{\left\langle \delta \mathbf{J\otimes }\delta \mathbf{J}\right\rangle}{2\delta t}
\end{equation}
 is a consequence of microreversibility~\citep{Rax2021a}.

\section{Angular momentum absorption}
\label{Sec:8}

Short of solving the quasilinear equation Eq.~(\ref{ql}), a clear understanding of the mechanism of angular momentum absorption (or emission) can be gained through the analysis of the quasilinear guiding center radial velocity $\left\langle \delta D\right\rangle /dt$ and quasilinear ICRH $\left\langle \delta J\right\rangle /dt$ identified in Eqs.~(\ref{einhjy}, \ref{einhjy5}). In general the effect of the wave is indeed two-fold: it both drives a radial drift current $\delta D$ and provides ICRH $\delta J$. Yet, this last channel should be avoided or at least minimized for fluid rotation sustainment. This can in principle be done through the choice of a suitable wave.

\subsection{Absorption from global angular momentum conservation}

For a single particle, the SAM ($S_{z}$) and OAM ($L_{z}$) lost by the wave during the resonant wave-particle interaction are gained by the particle in the form of canonical angular momentum $L_{C}$. From Eq.~(\ref{ccc})
\begin{equation}
-\frac{\left\langle \delta L_{C}\right\rangle }{\delta t} =\frac{%
\left\langle \delta J\right\rangle }{\delta t}-\frac{\left\langle \delta
D\right\rangle }{\delta t}
\end{equation}
which using Eq.~(\ref{einhjy5}) rewrites
\begin{equation}
-\frac{\left\langle \delta L_{C}\right\rangle }{\delta t} =\left( \sigma -n\right) \frac{\pi }{2}\sum_{\mathbf{N}}\mathbf{N}\cdot
\partial _{\mathbf{J}}\left[ \left| V_{\mathbf{N}}\right| ^{2}\delta \left( 
\mathbf{N}\cdot \bm{\Omega}-\omega \right) \right].  \label{angmo}
\end{equation}
Under the simple photon picture developed at the end of Sec.~\ref{Subec:resonance}, global angular momentum conservation for the full system wave plus particle and a single $\left| V_{\mathbf{N}}\right| $ coupling coefficient thus writes
\begin{equation}
\delta \left. L_{z}\right| _{wave}=-n\frac{\delta H_{0}}{\omega },\quad\delta \left. S_{z}\right| _{wave}=\sigma \frac{\delta H_{0}}{\omega }.  \label{saesae}
\end{equation}
Quasilinear theory brings additional insights in that it allows to relate the wave's change in angular momentum to the angular momentum absorption by a distribution function $F\left( \mathbf{J},t\right)$. Specifically, averaging Eq.~(\ref{angmo}) over the distribution of actions in the plasma and integrating by part gives
\begin{equation}
\left. \frac{d\left( L_{z}+S_{z}\right) }{dt}\right| _{wave}=\left( n-\sigma
\right) \frac{\pi }{2}\iint dJdD\sum_{l}\left| V_{\mathbf{N}}\right| ^{2}\mathbf{N}\cdot \partial _{\mathbf{J}}\left. F\right | _{P=P_{l}}.
\label{angmom2}
\end{equation}
Here $P_{l}$ is the resonant axial momentum fulfilling the
relation $\mathbf{N}\cdot \bm{\Omega}=\omega $ for given wave field and DC field configurations, that is
\begin{equation}
\ \beta P_{l}=\omega +\left( l+\sigma \right) \Omega _{+}-\left( l+n\right)
\Omega _{-}.  \label{re456}
\end{equation}
Note that since $-\infty <P<\infty $ there is always a solution $P_{l}$ to Eq.~(\ref{re456}) for a given $\left( l,n\right) \in \Bbb{Z}^{2}$. Note also that Eq.~(\ref{angmom2}) can be equivalently
derived from Eq.~(\ref{saesae}) using Eq.~(\ref{epr2}).

The angular momentum absorption coefficient derived in Eq.~(\ref{angmom2}) can be evaluated by considering the kinetic evolution of $F\left( \mathbf{J},t\right) $ given in Eq.~(\ref{ql}) together with a relaxation term associated with collisions. Assuming that the plasma equilibrium is only weakly perturbed by the wave we can consider that $F\sim F_{0}$ as given in Eq.~(\ref{max}).

\subsection{Physical picture}

To develop a deeper physical understanding of quasilinear angular momentum exchange we define the average kinetic angular momentum
\begin{equation}
\left\langle L_{K}\right\rangle  =\left\langle xv_{y}-yv_{x}\right\rangle
_{\theta +\varphi }=\frac{2\Omega _{-}}{\Omega }D+\frac{2\Omega _{+}}{\Omega 
}J  \label{def11}
\end{equation}
and the average magnetic flux through the orbit 
\begin{equation}
\left\langle \Psi \right\rangle  =\omega _{c}\pi \left\langle
x^{2}+y^{2}\right\rangle _{\theta +\varphi }=2\pi \frac{\omega _{c}}{\Omega }%
D+2\pi \frac{\omega _{c}}{\Omega }J.  \label{def22}
\end{equation}
According to the quasilinear prescription, the bracket $\left\langle{}\right\rangle $ indicates an angle average of Eqs.~(\ref{tri}, \ref{lll}, \ref{flux2}). Meanwhile, Eq.~(\ref{flux mom}) gives a relation for the
canonical angular momentum variation 
\begin{equation}
\delta L_{C}=\delta \left\langle L_{K}\right\rangle +\delta \left\langle
\Psi \right\rangle /2\pi .  \label{kkll2}
\end{equation}
From Eq.~(\ref{def11}) and Eq.~(\ref{def22}) the kinetic and magnetic components $\delta \left\langle L_{K}\right\rangle$ and $\delta\left\langle \Psi \right\rangle$ then write
\begin{gather}
\Omega \delta \left\langle L_{K}\right\rangle  =2\Omega _{-}\delta
D+2\Omega _{+}\delta J,  \label{kkll} \\
\Omega \delta \left\langle \Psi \right\rangle  =2\pi \omega _{c}\delta
D+2\pi \omega _{c}\delta J.
\end{gather}
A physical interpretation of these results can be obtained as follows, where we focus again on the more intuitive ordering $D>J$.

Starting with the kinetic component Eq.~(\ref{kkll}), recall from Sec.~\ref{Sec:4} that an ion with mass $M=1$ displays a guiding center moment of inertia $M_{G}$ = $2D/\Omega $ with respect to the $z$ axis, and a moment of inertia of the cyclotron motion $M_{C}$ = $2J/\Omega $ with respect to the guiding center. Recall also that the guiding center of this ion rotates at $d\theta /dt = \Omega _{-}$ whereas the cyclotron rotation takes place at the angular frequency $\Omega _{+} = -d\varphi /dt$. Now, because these two angular velocities are set by the fields Eqs.~(\ref{chan1}, \ref{chan2}), the variation of the kinetic angular momentum of the ion $\left\langle L_{K}\right\rangle = M_{G}\Omega _{-} + M_{C}\Omega _{+}$ ($\theta$ is anticlockwise and $\varphi$ is clockwise) must come from a variation of the moments of inertia and not of the angular
velocities, so that
\begin{equation}
\frac{\delta \left\langle L_{K}\right\rangle }{\delta t}=\frac{%
\delta M_{G}}{\delta t}\Omega _{-}+\frac{\delta M_{C}}{\delta t}\Omega _{+}=%
\frac{2\Omega _{-}}{\Omega }\frac{\delta D}{\delta t}+\frac{2\Omega
_{+}}{\Omega }\frac{\delta J}{\delta t},  \label{c1}
\end{equation}
which is precisely Eq.~(\ref{kkll}). 

The interpretation of the magnetic component
\begin{equation}
\frac{1}{2\pi }\frac{\delta \left\langle \Psi \right\rangle }{\delta t}=\frac{\omega _{c}}{\Omega }\frac{\delta D}{\delta t}+\frac{\omega _{c}}{\Omega }\frac{\delta J}{\delta t},  \label{c2}
\end{equation}
requires an analysis of both the $\delta D$ and $\delta J$ terms. As we will now show, these two terms can be interpreted in terms of two different torques exerted on an ion in the background magnetic field. Starting with $D$, two pictures can be invoked. The first one is to consider the axial torque due to the magnetic force exerted on a charge $q=1$ moving radially. The radial velocity of this charge is 
\begin{equation}
 \frac{d}{dt}\left(\sqrt{\frac{2D}{\Omega}}\right) \mathbf{e}_{r},
\end{equation}
so that this torque writes
\begin{equation}
\sqrt{\frac{2D}{\Omega }}\mathbf{e}_{r}\times \left[\frac{d}{dt}\left(\sqrt{\frac{2D}{\Omega}}\right)\mathbf{e}_{r}\times \omega _{c}\mathbf{e}_{z}\right] =-\frac{\omega
_{c}}{\Omega }\frac{dD}{dt}\mathbf{e}_{z}.
\end{equation}
The second is to consider an azimuthal electromotive force (emf) for a $q=1$ charge distributed along a rotating circle with radius $\sqrt{2D/\Omega}$, associated with the variation of the loop surface as a result of the radial motion.  This electric inductive field $E_{emf}$ is also the source of an axial torque 
\begin{equation}
\sqrt{\frac{2D}{\Omega }}\mathbf{e}_{r}\times E_{emf}\mathbf{e}_{\alpha }=-\frac{1}{2\pi }\frac{d\Psi _{G}}{dt}\mathbf{e}_{z}=-\frac{\omega _{c}}{\Omega }\frac{dD}{dt}\mathbf{e}_{z},
\end{equation}
where we have introduced the magnetic flux through the guiding center orbit $\Psi _{G}$ and applied Faraday's law 
\begin{equation}
-\frac{d\Psi _{G}}{dt}=\oint E_{emf}\mathbf{e}_{\alpha }\mathbf{\cdot }ds%
\mathbf{e}_{\alpha }=2\pi \sqrt{\frac{2D}{\Omega }}E_{emf}.
\end{equation}
One verifies that both analyses give the same axial torque experienced by an ion as a result of the wave driven radial motion, which is precisely the first term on the right hand side in Eq.~(\ref{c2}). Moving on finally to the $\delta J$ term in Eq.~(\ref{c2}), a similar current loop picture can be brought up but by considering a $q=1$ charge distributed this time along the Larmor radius $\sqrt{2J/\Omega}$. The magnetic flux through this varying Larmor radius in indeed $\Psi_{L}=2\pi\left(\omega _{c}/\Omega \right) J$, whose time derivative precisely gives back the second term on the right hand side in Eq.~(\ref{c2})

In summary, the first term on the right hand side of Eq.~(\ref{kkll2}) corresponds to a change of the moment of inertia of the particle as a result of the quasilinear radial drift and Larmor radius evolution. The second term on the right hand side of Eq.~(\ref{kkll2}) corresponds to torques which result from a change in magnetic fluxes. Both of these terms, interpreted here in the weak field limit, must be balanced by the transfer of a corresponding angular momentum from the wave. Note however that the wave quasilinear primary effect is not a torque, it is a wave driven radial current.

\section{Radial current generation}
\label{Sec:9}



Two types of model can be constructed from the quasilinear kinetic equation derived in Eq.~(\ref{ql}). One option is to balance the quasilinear evolution of the distribution function with a collisional evolution at the
kinetic level and then to average the solution $F\left( \mathbf{J}\right) $ to obtain a steady state fluid picture. The other is to average the quasilinear dynamics to derive the fluid quasilinear flows of mass, charge and momentum, and then to balance these fluid flows with the dissipative terms involved in classical transport theory. The latter option is used in the following, where we further assume for simplicity axial homogeneity along $z$. 

Let us write
\begin{equation}
\bm{\Gamma }\left( r\right)  = \Gamma _{r}\mathbf{e}_{r}+\Gamma_{\alpha }\mathbf{e}_{\alpha }+\Gamma _{z}\mathbf{e}_{z}
\end{equation}
the wave driven particle flux which results from the absorption of the wave power $w_{\mathbf{N}}\left( r\right) $ at radius $r$. The radial flow $\Gamma _{r}$ is due to the wave driven guiding center radial velocity, which we showed is proportional to $\left\langle \delta D\right\rangle /\delta t$. The azimuthal flow $\Gamma _{\alpha }$ is a small diamagnetic effect associated with inhomogeneous wave driven ICRH proportional to $\left\langle
\delta J\right\rangle /\delta t$. The axial flow $\Gamma _{z}$ is simply the wave driven current from classical current generation~\citep{Fisch1978,Fisch1987} due to $\left\langle \delta P\right\rangle /\delta t$. The amplitude of these fluxes is governed by the evolution equation Eq.~(\ref{einhjy}).

An analytical expression for the radial flow $\Gamma _{r}$ can be derived if focusing once again on the familiar limit $J<D$. In this limit the average radial position of a resonant particle is $r^{2}=2D/\Omega $, and the average Lagrangian radial velocity is hence $r\Omega dr/dt$ = $dD/dt$. From Eq.~(\ref{einhjy}) the phase averaged evolution of $D$ is proportional to the absorbed power $w_{\mathbf{N}}$ associated with the $l$ harmonic Brillouin resonance for a rotating wave with azimuthal number $n$, and one has
\begin{equation}
\left. \frac{\left\langle \delta D\right\rangle }{\delta t}\right| _{\mathbf{N}}=\frac{l+n}{\omega }\left. w_{\mathbf{N}}\right| _{D=\frac{\Omega r^{2}}{2}}.
\end{equation}
Introducing back the particle mass $M$, and using the power density absorbed by the plasma $W_{\mathbf{N}}$ (in Watt/m$^{3}$) rather than $w_{\mathbf{N}}$ (in Watt), the wave driven
resonant particle flux is 
\begin{equation}
\Gamma _{r}=\frac{\left( l+n\right) }{rM\Omega \omega }W_{\mathbf{N}} \quad\textrm{s}^{-1}.~\textrm{m}^{-2}.
\label{c3}
\end{equation}

Consider now that the wave power density $W_{\mathbf{N}}\left( t\right)$ is turned on adiabatically at $t$ = $-\infty $ with $W_{\mathbf{N}}\left( -\infty\right) =0$. The wave moves some resonant particles across the magnetic field, which leads to a radial current $J_{W}\left( t\right)$ such that $J_{W}\left( t=-\infty \right) = 0$ and $J_{W}\left( t=0\right) = q\Gamma _{r}$. The resulting time evolution of the radial electric field can be described as follows. From an electrical point of view, the build-up corresponds to a capacitive electric
field build up in a dielectric media, akin to the charging of a cylindrical capacitor. By considering the plasma as a dielectric with low frequency permittivity
\begin{equation}
\varepsilon = 1 + \left(\frac{\omega _{pi}}{\omega}\right)^{2} \approx \left(\frac{\omega _{pi}}{\omega}\right)^{2}
\end{equation}
with $\omega _{pi}$ the ion plasma frequency, the electric field $\mathbf{E}\left( t\right)$ throughout this transient phase is thus determined from Maxwell-Amp\`{e}re equation
\begin{equation}
\varepsilon _{0}\frac{\omega _{pi}^{2}}{\omega _{ci}^{2}}\frac{\partial 
\mathbf{E}}{\partial t}+J_{W}\left( t\right)\mathbf{e}_{r} =\mathbf{0}.
\label{maxw}
\end{equation}
From a mechanical point of view this build-up phase corresponds to an angular momentum input via the $J_{W}\left( t\right)\mathbf{e}_{r} \times \mathbf{B}$ force, and this momentum is converted into plasma $E\times B$ drift. Indeed, integrating Eq.~(\ref{maxw}) over the
transient phase gives
\begin{equation}
\int_{-\infty }^{0}J_{W}\left( t\right)\mathbf{e}_{r} \times \mathbf{B}dt=-N_{p}M\frac{\mathbf{E}_{0}\times \mathbf{B}}{B^{2}}
\end{equation}
with $\mathbf{E}_{0}=\mathbf{E}\left( t=0\right)$, $M$ the ion mass and $N_{p}$ the ion density, which confirms this momentum balance. Note that ion diamagnetic effects have been neglected in writing Eq.~(\ref{maxw}). Note also that because the conductivity along magnetic field lines is generally far larger than the conductivity across the field lines, charges rapidly move away from the wave active regions along the field lines, which in turn become equipotential. 

Finally, for $t>0$, the charge separation induced by the wave is short circuited by the plasma perpendicular conductivity~\citep{Helander2005, Rax2019,Kolmes2019}. In this steady-state dissipative regime the wave driven current $q\Gamma _{r}\mathbf{e}_{r}$ is balanced by a weak discharging Ohmic current $\mathbf{J}_{\text{conduction}}$, with 
\begin{equation}
\mathbf{\nabla }\cdot \left[ q\Gamma _{r}\mathbf{e}_{r}+\mathbf{J}_{\text{%
conduction}}\right] =0.  \label{divdiv}
\end{equation}
Thus, as opposed to axial current generation, the steady-state is determined by the geometry of the plasma. The examination of this problem is left for a future study.

\section{Conclusions}
\label{Sec:10}

Although angular momentum exchange between a wave and a rotating plasma is of importance both to astrophysics~\citep{Goldreich1969,Julian1973,Ferriere2006} and laboratory plasmas~\citep{Shvets2002,Kostyukov2002,Thaury2013}, a kinetic model of this interaction had to our knowledge never been proposed. In this study we addressed this issue and derived the quasilinear equation describing the interaction between a rotating wave and a rotating magnetized plasma. We further used this kinetic model to analyze angular momentum absorption/emission and to understand the interplay between orbital angular momentum (OAM), spin angular momentum (SAM) and finite Larmor radius (FLR) effects.


First, a canonical angle-action Hamiltonian description of the unperturbed Brillouin rotation dynamics in a rotating plasma is derived. The identification of angle-action variables allows to separate the fast part of the unperturbed motion (angles) from the constant (integrable system) or slow (adiabatic system) part of the unperturbed motion (actions). Through this process we identify three canonical actions $D$, $J$ and $P$. The latter is the classical momentum along $\mathbf{B}$. The first two $D$ and $J$ are shown in the weak field limit, that is, for a cross-field drift frequency $E/(rB)$ small compared to the ion gyro-frequency, to be related to the magnetic flux through the cyclotron orbit and the guiding center orbit.

Then, the wave-particle coupling is expressed in terms of these angle-action variables in the form of a perturbed Hamiltonian. This approach made it possible to identify a new resonance condition, which generalizes the classical Landau-cyclotron resonance to the case of a rotating plasma interacting with a rotating wave. This new condition, Eq.~(\ref{res 65}), is referred to as Brillouin resonance. It notably expresses finite Larmor radius effects through Eqs.~(\ref{ph1}, \ref{ph2}, \ref{ph3}), which are found to be responsible for the occurrence of harmonic ($l$) Brillouin resonances. Together with this resonance condition, diffusion paths in action space were identified, and particular examples were exposed in the weak field limit. Finally, the quasilinear equation Eq.~(\ref{ql}) which describes energy-momentum exchange as a time evolution of the actions distribution function was derived by averaging the kinetic response of the plasma to the perturbation over the fast part of the motion (angles).

By analyzing the variation in canonical actions $D$ and $J$ predicted by the quasilinear equation, a physical picture for momentum absorption was finally proposed. Specifically, angular momentum from the wave was shown to be transferred to the plasma either as a change of the inertia tensor of the plasma, or as a magnetic flux variation Eqs.~(\ref{c1},~\ref{c2}). This analysis also showed that the radial flux can be identified as the source of angular momentum injection in the plasma. 
An interesting prospect is the generalisation of this work to magnetic field inhomogeneities. The use of action-angle coordinates for the motion in a straight magnetic field with constant gradient~\citep{Brizard2022} may for instance enable to capture bounce resonances in a mirror geometry rather than the uniform $z$ translation Doppler shift considered here.

Finally, since the radial current is proportional to the absorbed power, sustaining steady-state rotation with waves will require adjusting the wave power deposition profile in a way that Eq. (\ref{divdiv}) is fulfilled. The optimization of power deposition will be the object of future studies, but we note for example that for a resonance $\mathbf{N}$ in a rotating magnetized plasma cylinder with large conductivity along the field lines $\eta_{\shortparallel }$ and weak conductivity across the field lines $\eta_{\perp }$, Eq.~(\ref{divdiv}) gives 
\begin{equation}
W_{\mathbf{N}}\left( r\right) \approx \eta _{\perp }\frac{M^{2}}{q^{2}}\frac{\Omega \omega }{\left( l+n\right) }\left( \frac{\omega _{c}^{2}-\Omega ^{2}}{4}\right) r^{2}.
\end{equation}
How this power density can be deposited in the plasma requires information on what the normal modes of a rotating plasma column are, and in particular the dispersion relation characterising these modes, beyond the simple case of an aligned rotator~\citep{Gueroult2019a,Gueroult2020,Gueroult2023}. This important question will be addressed in forthcoming studies.

\section*{Acknowledgments}

The authors would like to thank Dr. I. E. Ochs, E. J. Kolmes, T. Rubin, and M. E. Mlodik for constructive discussions. This work was supported by grants DOE DE-SC0016072, NNSA DE-SC0021248 and ANR-21-CE30-0002 (Project WaRP). JMR acknowledges Princeton University and the Andlinger Center for Energy + the Environment for the ACEE fellowship which made this work possible.

\appendix

\section{SAM and OAM of a vector field}
\label{Sec:SAMOAM}

Consider a wave field $\mathbf{A}\left( \mathbf{r}\right) \exp j\omega t$. The identification of (\textit{i}) linear momentum, (\textit{ii}) spin and (\textit{iii}) orbital angular momentum eigenstates can be guided by the analysis of the transformation properties of the wave under translations and rotations. 

For this consider first the change of this vector field $\mathbf{A}\left( \mathbf{r}\right) $ under an active (change of the object $\mathbf{A}$), or a passive (change of the frame and coordinates used to describe $\mathbf{A}$), infinitesimal translation $\widehat{T}$ associated with the small vector $\delta \mathbf{r}$\begin{equation}
\widehat{T}\mathbf{A}\left( \mathbf{r}\right) =\mathbf{A}\left( \mathbf{r}\pm \delta \mathbf{r}\right) =\mathbf{A}\left( \mathbf{r}\right) \pm \delta 
\mathbf{r.\nabla A}\left( \mathbf{r}\right) .  \label{geo}
\end{equation}
The minus or plus signs are associated with the passive or the active points of view. Eq.~(\ref{geo}) can be rewritten as a near identity transformation 
\begin{equation}
\widehat{T}\mathbf{A}\left( \mathbf{r}\right) =\left[ \mathbf{I}\pm j\delta\mathbf{r.}\widehat{\mathbf{P}}\right] \mathbf{A}\left( \mathbf{r}\right),
\end{equation}
where $\mathbf{I}$ is the identity operator. The linear momentum operator $\widehat{\mathbf{P}}$ is defined in the usual way as $\widehat{\mathbf{P}}=-j\mathbf{\nabla }$. The eigenvectors of this linear momentum operator $\widehat{\mathbf{P}}$ \ are the plane waves, $\widehat{P_{z}}\left( \exp j\beta z\right) $ = $\beta \left( \exp j\beta z\right) $, which are also solutions of Maxwell-Amp\`{e}re and Maxwell-Faraday equations in an homogeneous linear dispersive plasma provided that $\beta\left(\omega \right) $ fulfills the dispersion relation.

Consider now the change of a vector field $\mathbf{A}\left( \mathbf{r}\right) $ under an active, or a passive, infinitesimal rotation $\widehat{R}$ associated with a small $\delta \alpha $ turn around an axis directed by a
given unit vector $\mathbf{n}$ ($\mathbf{n}^{2}=1$) 
\begin{align}
\widehat{R}\mathbf{A}\left( \mathbf{r}\right) & =\pm \delta \alpha \mathbf{n}\times \mathbf{A}\left( \mathbf{r}\pm \delta \alpha \mathbf{n}\times \mathbf{r}\right)\nonumber \\
& =\mathbf{A}\left( \mathbf{r}\right) \pm \delta \alpha \left[ 
\mathbf{n}\times +\left( \mathbf{n}\times \mathbf{r}\right) \mathbf{.\nabla }\right] \mathbf{A}\left( \mathbf{r}\right) .  \label{geo2}
\end{align}
Eq.~(\ref{geo2}) can be rewritten as a near identity transformation displaying the separation between OAM and SAM operators 
\begin{equation}
\widehat{R}\mathbf{A}\left( \mathbf{r}\right) =\left[ \mathbf{I}\pm j\delta
\alpha \mathbf{n\cdot }\left( \widehat{\mathbf{L}}+\widehat{\mathbf{S}}\right) \right] \mathbf{A}\left( \mathbf{r}\right) .
\end{equation}
The OAM operator $\widehat{\mathbf{L}}$ \ and the SAM operator $\widehat{\mathbf{S}}$ are thus defined according to the usual relations 
\begin{gather}
\widehat{\mathbf{L}} =-j\mathbf{r}\times \mathbf{\nabla } \\
\widehat{\mathbf{S}} =-j\mathbf{n}\times
\end{gather}
Note that, since $\widehat{S}_{x}^{2}+\widehat{S}_{y}^{2}+\widehat{S}_{z}^{2}=2\mathbf{I}$, we recover the usual angular momentum rule for a vector : $\widehat{\mathbf{S}}^{2} = s\left( s+1\right) \mathbf{I}$ with $s = 1 $. 

We now restrict the transformations to rotations around the magnetic field direction in which case the angular momentum operator reduces to its $z$ component $\widehat{L}_{z}+\widehat{S}_{z}$. The eigenvectors of the (\textit{i}) OAM operator $\widehat{L}_{z}$ and of the (\textit{ii}) SAM operator\ $\widehat{S}_{z}$ are (\textit{i}) $\exp \pm jn\alpha $, where $\alpha $ is the polar angle around the magnetic field, and (\textit{ii}) the L and R circularly polarized waves basis with eigenvalues $\pm 1$ and $\mathbf{e}_{z}$ with
zero eigenvalue 
\begin{equation}
\left( \widehat{L}_{z}+\widehat{S}_{z}\right) \left( \frac{\mathbf{e}%
_{x}\pm j\mathbf{e}_{y}}{\sqrt{2}}\exp jn\alpha \right) =\left( n\mp
1\right) \left( \frac{\mathbf{e}_{x}\pm j\mathbf{e}_{y}}{\sqrt{2}}\exp
jn\alpha \right) .  \label{s3}
\end{equation}
Solutions of Maxwell-Amp\`{e}re and Maxwell-Faraday equations with a factor $%
\exp \pm jn\alpha $ have a well defined OAM in rotating magnetized plasma
when the magnetic axis is also the rotation axis. Solution of
Maxwell-Amp\`{e}re and Maxwell-Faraday equations with a polarization $%
\mathbf{e}_{x}\pm j\mathbf{e}_{y}$ have a well defined SAM.

\section{Canonical quasilinear equation}
\label{Sec:AppendixB}

In this appendix we briefly review the derivation of the canonical quasilinear equation~\citep{Rax2021a}. 

Consider an integrable Hamiltonian $H_{0}$ (the adiabatic trap) and an oscillating perturbation (the wave) such that $V\ll H_{0}$. The kinetic description of wave particle interaction can be performed through a two time scales separation : $F\left( \mathbf{J},t\right) $ is the distribution function in action space describing the slow evolution ($\partial _{t}F\left( \mathbf{J},t\right) \sim O\left( V^{2}\right) $) of a given population and $f\left( \mathbf{J},\bm{\theta },t\right) \sim O\left( V\right) $ describes the fast oscillating evolution ($\partial
_{t}f\left( \mathbf{J},\bm{\theta },t\right) \sim O\left( V\right) $) in phase space
\begin{gather}
\mathcal{H} =H_{0}\left( \mathbf{J}\right) +V\left( \mathbf{J},\mathbf{\theta },t\right) , \\
\mathcal{F} =F\left( \mathbf{J},t\right) +f\left( \mathbf{J},\mathbf{\theta },t\right) .
\end{gather}
Here $\left( \mathbf{J},\bm{\theta }\right) $ is a set of action-angle variables for the unperturbed dynamics, $\bm{\Omega }=\partial H_{0}/\partial \mathbf{J}$, $\mathcal{H}$ is the perturbed Hamiltonian and $\mathcal{F}$ is the distribution function providing a kinetic description of the perturbed dynamics. Liouville's equation can be written with the help of Poisson bracket as $\partial _{t}\mathcal{F}$ = $\left\{ \mathcal{H},%
\mathcal{F}\right\} $
\begin{equation}
\frac{\partial }{\partial t}\left( F+f\right) =\left\{ \left( H_{0}+V\right),\left( F+f\right) \right\}  \label{lili}
\end{equation}
Because $\left\{ H_{0},F\right\} =0$ we can split this relation into (\textit{i}) an $O\left( V\right) $ fast evolution - the Vlasov equation Eq.~(\ref{vvv12}) - and (\textit{ii}) an $O\left( V^{2}\right) $ slow secular
evolution - the quasilinear equation Eq. (\ref{ql876}) - , 
\begin{gather}
\text{{}}\frac{\partial f}{\partial t} =\left\{ V,F\right\} +\left\{
H_{0},f\right\} \sim O\left( V\right) ,  \label{vvv12} \\
\text{{}}\frac{\partial F}{\partial t} =\left\langle \left\{ V,f\right\}
\right\rangle _{\bm{\theta }}\sim O\left( V^{2}\right) ,
\label{ql876}
\end{gather}
where we write $\left\langle {}\right\rangle _{\bm{\theta }}$ the average
over the fast rotating angles $\bm{\theta }$.

The next step is to consider a Fourier decomposition of the $O\left(V\right) $ oscillating Vlasov terms. This decomposition is always possible as $f$ and $V$ are periodic functions of the angle $\bm{\theta }$,
through the derivation of the Fourier coefficient often requires some lengthy calculations
\begin{gather}
V\left( \mathbf{J},\bm{\theta },t\right)  =\sum_{\mathbf{N}}V_{\mathbf{N}}\left( \mathbf{J}\right) \exp j\left( \mathbf{N}\cdot \bm{\theta }-\omega t\right)   \label{svfl2} \\
f\left( \mathbf{J},\bm{\theta },t\right)  =\sum_{\mathbf{N}}f_{\mathbf{N}}\left( \mathbf{J}\right) \exp j\left( \mathbf{N}\cdot \bm{\theta }-\omega t\right) 
\end{gather}
where $N\in \Bbb{Z}^{3}$. With this Fourier decomposition Vlasov's equation
Eq. (\ref{vvv12}) becomes an algebraic equation whose solution is 
\begin{equation}
\text{{}}f=\sum_{\mathbf{N}}\frac{V_{\mathbf{N}}\left( \mathbf{J}\right) }{\mathbf{N}\cdot \bm{\Omega} -\omega }\mathbf{N}\cdot \frac{\partial F}{\partial \mathbf{J}}+j\pi \sum_{\mathbf{n}}V_{\mathbf{N}}\left( \mathbf{J}\right) \delta \left( \mathbf{N}\cdot \bm{\Omega}-\omega \right) \mathbf{N}\cdot \frac{\partial F}{\partial \mathbf{J}}.
\end{equation}
We recognize the adiabatic part of the plasma response, which ultimately provides the Hermitian part of the dielectric tensor, and the resonant part which provides the description of collisionless dissipation. In order to
average Eq.~(\ref{ql876}) 
\begin{equation}
\frac{\partial F}{\partial t}=\left\langle \frac{\partial V}{\partial \bm{\theta }}\cdot \frac{\partial f}{\partial \mathbf{J}}-\frac{\partial f}{\partial \bm{\theta }}\cdot \frac{\partial V}{\partial 
\mathbf{J}}\right\rangle _{\bm{\theta }}
\end{equation}
we use the usual rule $\left\langle \Real[a\left( u\right)] \Real[b\left( u\right)] \right\rangle _{u}=\Real\left[ a\left( u\right)b^{*}\left( u\right) \right] /2$ and finally obtain the canonical form 
\begin{equation}
\frac{\partial F}{\partial t}=\partial_{\mathbf{J}}\cdot \sum_{\mathbf{N}}\mathbf{N}~\Real\left( j\frac{V_{\mathbf{N}}f_{\mathbf{N}}^{*}}{2}\right) =\frac{\pi }{2}\frac{\partial }{\partial \mathbf{J}}\mathbf{%
\cdot }\left[ \sum_{\mathbf{N}}\mathbf{N}\left| V_{\mathbf{N}}\right|^{2}\delta \left( \mathbf{N}\cdot \bm{\Omega}-\omega \right) \mathbf{N}\cdot \frac{\partial F}{\partial \mathbf{J}}\right].   \label{ql568}
\end{equation}
used in Eq. (\ref{ql}).

\bibliographystyle{jpp}

\bibliography{JMR_QuasiLinear}

\end{document}